\documentclass[letterpaper,showpacs,aps]{revtex4-2}

\usepackage[utf8]{inputenc}
\usepackage[english]{babel}

\usepackage{latexsym}
\usepackage{amssymb}
\usepackage{amsmath}
\usepackage{bm} 

\usepackage{epsfig}
\usepackage{graphicx}
\usepackage{float}

\usepackage{color}

\begin{document}


\title{Charged boson stars revisited}

\author{José Damián López}
\email{damian13.03@ciencias.unam.mx}

\author{Miguel Alcubierre}
\email{malcubi@nucleares.unam.mx}

\affiliation{Instituto de Ciencias Nucleares, Universidad Nacional
Aut\'onoma de M\'exico, A.P. 70-543, M\'exico D.F. 04510, M\'exico.}


\date{\today}


\begin{abstract}
We consider again stationary solutions to the spherically symmetric
Einstein--Maxwell--Klein--Gordon system, commonly known as ``charged
boson stars'', originally studied by Jetzer and Van Der Bij
in~\cite{Jetzer89}. We construct families of charged boson stars in
the ground state, for different values of the charge parameter $q$,
and different values of the central scalar field. Following Jetzer and
Van Der Bij, one can define a critical value for the charge $q=q_c$
that corresponds to the value for which the Coulomb repulsion of the
bosonic particles exactly cancels their newtonian gravitational
attraction. We confirm the claim made by Pugliese et
al. in~\cite{Pugliese_2013} that super-critical solutions exist for a
limited range of charges above the critical value $q>q_c$ (though we
find an even smaller range of $q$ for which this is possible).  Our
analysis indicates, however, that all such super-critical solutions
are gravitationally unbound, and are therefore expected to be
unstable. One of the main results of our analysis is the fact that,
even though we do find a family of slightly super-critical solutions
in the sense that $q>q_c$, there are {\em no}\/ super-critical
solutions in the sense that the total charge $Q$ is larger than the
total mass $M$ of the system.
\end{abstract}


\pacs{
04.20.Ex, 
04.25.Dm, 
95.30.Sf  
}


\maketitle


\section{Introduction}
\label{sec:intro}

Boson stars are self-gravitating soliton-type configurations for a
massive complex scalar field in general relativity
(see~\cite{Jetzer92,Lee1992251,Liebling:2012fv,Visinelli_2021,Schunck:2003kk}
for reviews). These compact objects have been extensively studied
since the pioneering works of Kaup in 1968~\cite{Kaup68}, where they
are referred to as a ``Klein--Gordon Geon'', and of Ruffini and
Bonazzola in 1969~\cite{Ruffini69}. In 1987, the same compact object
was studied by Friedberg, Lee, and Pang~\cite{Friedberg87}, where it
was called a ``mini-soliton star''. Boson Stars can be considered
descendants of the self-gravitating photonic configurations called
geons (gravitational electromagnetic units) proposed by Wheeler in
1955~\cite{Wheeler1955}. Standard boson stars (also known as
mini-boson stars) are described by a massive complex scalar field that
is localized in a compact spatial region and is supported by its
self-gravity. Complex scalar fields allow compatibility with a static
geometry of spacetime, such that the scalar field contains two degrees
of freedom, both oscillating harmonically in time, but out of
phase. In this way, it is possible to evade the no-soliton Derrick
Theorem~\cite{Derrick1964}. The case of real scalar fields has also
been studied (see e.g.~\cite{Seidel_PhysRevLett.66.1659}), though in
this case one can only form quasi-stationary configurations known as
``oscillatons''.

Although boson stars remain as purely theoretical, interest in these
self-gravitating compact objects has recently increased due to
developments in both particle physics and cosmology, such as the
confirmation of the Higgs boson~\cite{Aad:2012tfa}, suggesting that in
the early stages of the universe boson stars may have formed out of
fundamental scalar fields and could play a role in understanding the
origin of dark matter. Furthermore, it has been shown that boson stars
can be considered as candidates for black hole
mimickers~\cite{Torres2000,Guzman:2005bs}. Different types of bosonic
structures have also been studied, such as the so-called ``Proca
stars'', which are analogous to boson stars but for the case of a
massive complex abelian vector field minimally coupled to gravity. In
a similar way to the scalar field, the massive complex abelian vector
field can also form self-gravitating solutions to the Einstein--Proca
system~\cite{Sanchis_Gual_2017,Obukhov_1999,BRITO2016291}. Another
class of static solutions to the Einstein--Klein--Gordon equations
recently studied are the $\ell$-boson stars, which incorporate the
effects of angular momentum while maintaining the spherical symmetry
of the spacetime~\cite{Alcubierre:2018ahf,Alcubierre_2022}.

In this work we focus on the case of charged boson stars, which are spherically symmetric self-gravitating solutions for a massive complex scalar field coupled to the gauge group $U(1)$: the Einstein--Maxwell--Klein--Gordon (EMKG) system. Charged boson stars were first studied by Jetzer and Van Der Bij in~\cite{Jetzer89}, and later by Pugliese et al. in~\cite{Pugliese_2013}. More recently, the EMKG system has also been studied in different contexts, such as the case of fermion-charged-boson stars in~\cite{Kain_2021}, or the gravitational collapse of charged matter in~\cite{Torres_2014}. Charged boson stars can be characterized by the charge parameter $q$ associated to the bosonic conserved current. In~\cite{Jetzer89} Jetzer and Van Der Bij argue that there is a critical value for the charge $q_{c}$ such that charged boson stars can only exist for $q<q_{c}$. In the original units of~\cite{Jetzer89} (see discussion at the end of Section~\ref{sec:matter} below), the critical value is given by $q_c = 1/\sqrt{2} \: (m/M_{PL}) \simeq 0.707 \: (m/M_{PL})$, with $m$ the mass parameter of the complex scalar field and $M_{PL}=\sqrt{\hbar c/G}$ the Planck mass. In newtonian terms, this critical value for the charge corresponds to the case when the Coulomb repulsion exactly cancels the gravitational attraction of the fundamental bosonic particles, so that for super-critical values of the charge one would not expect to find stationary solutions.  However, it was claimed in~\cite{Pugliese_2013} that solutions with super-critical charge up to $q \sim 0.8 \: (m/M_{PL}) > q_c$ can in fact be found.

In this work we study again in detail the solutions for charged boson stars, and show that indeed there exits a family of solutions corresponding to charged boson stars with a super-critical charge $q>q_{c}$. We find, however, that some of the super-critical solutions of Pugliese et al. are not correct, apparently due to the fact that the spatial range they considered did not extend far enough, so that when the spatial domain is extended the scalar field does not decay exponentially and the resulting space-time is not asymptotically flat. In our case we can only find super-critical solutions up to  $q \sim 0.739 \: (m/M_{PL})$.  However, one of our main results is the fact that, even though we do find a family of slightly super-critical solutions in the sense that $q>q_c$, there are no super-critical solutions
in the sense that the total charge $Q$ is larger than the total mass $M$ of the system (see Section~\ref{sec:results} below).

In order to solve the EMKG system we use a 3+1 decomposition of the electromagnetic field as described in~\cite{Torres_2014,Alcubierre:2009ij}, and construct families of stationary solutions for different
values of the charge parameter $q$.  For a given value of $q$, the system of equations can then be cast as an eigenvalue problem for the oscillation frequency of the scalar field $\omega$, with the value of the scalar field at the origin as a free parameter. We have also performed short time evolutions of some of our solutions in order to verify that the frequencies obtained in our analysis do correspond to the frequencies observed during a dynamical evolution. For these evolutions we have used a fully non-linear time evolution code for numerical relativity in spherical symmetry, the OllinSphere code previously described in~\cite{Torres_2014,Alcubierre:2010ea,Ruiz:2012jt}, that uses the Baumgarte--Shapiro--Shibata--Nakamura formulation adapted to spherical symmetry~\cite{Shibata95,Baumgarte:1998te,Brown:2009dd,Alcubierre:2010is}. However, we will leave a full analysis of these time evolutions for a future work.

This paper is organized as follows. In Section~\ref{sec:matter} we present the Einstein--Maxwell--Klein--Gordon system, and also discuss the units conventions taken for the electromagnetic field, as well as the normalization that we use for the charge parameter $q$. In Section~\ref{sec:bosonstar} we derive the spherically symmetric field equations for stationary configurations corresponding to the charged boson stars. In Section~\ref{sec:mass-charge} we define the total mass, total charge, and binding energy of the system. Section~\ref{sec:boundaries} describes the boundary conditions and the numerical methods that we have used to ensure that the solutions obtained are asymptotically flat. Section~\ref{sec:results} presents our results for the different families of solutions.  We conclude in Section~\ref{sec:CONCLUSIONS}.


\section{The Einstein--Maxwell--Klein--Gordon system}
\label{sec:matter} 

A self-gravitating charged massive complex scalar field $\phi$ is described by the following action (we use a metric signature $(-,+,+,+)$ and Plank units such that $G=c=\hbar =1$):
\begin{equation}
    S = \int  \left( \frac{R}{16\pi}
- \frac{1}{2} \left[ \left( \mathcal{D}_{\mu} \phi \right)^{*}
\left( \mathcal{D}^{\mu} \phi \right) + m^2 |\phi|^2 \right] - \frac{1}{16 \pi} \: \mathcal{F}^{\mu\nu} \mathcal{F}_{\mu\nu} \right)
\sqrt{-g} \: dx^4 \; ,
\label{actionTOTAL}
\end{equation}
where $g$ is the determinant of the space--time metric, $R$ its associated Ricci scalar, $m$ is the scalar field mass parameter, $\mathcal{F}_{\mu\nu}$ is the electromagnetic Faraday field tensor (2-form) given in terms of the potential 1-form $A_\mu$ as:
\begin{equation}
\mathcal{F}_{\mu\nu} := \partial_{\mu} A_{\nu} - \partial_{\nu} A_{\mu} \; ,
\label{DefsFaradayGaugeDev2}
\end{equation}
and $\mathcal{D}_{\mu}$ is the gauge invariant covariant derivative:
\begin{equation}
\label{DefsGaugeDev}
\mathcal{D}_{\mu} := \nabla_{\mu} + iqA_{\mu} \; ,
\end{equation}
with $\nabla_\mu$ the usual space--time covariant derivative and $q$ the scalar field charge parameter.

The action (\ref{actionTOTAL}) is invariant under local $U(1)$ gauge transformations of the form:
\begin{equation}
\label{GaugeTrans}
\phi \rightarrow e^{i q \theta(x^{\alpha})} \phi \; , \qquad
A_\mu \rightarrow A_\mu - \partial_\mu \theta(x^{\alpha}) \; ,
\end{equation}
with $\theta(x^{\alpha})$ a local gauge function which depends on the space--time coordinates. This implies the existence of a conserved Noether 4-current, which acts as a source for the electromagnetic field and is given by:
\begin{eqnarray}
\label{4Current}
j_\mu &=& \frac{iq}{2} \left[ \phi^{*} \mathcal{D}_{\mu}\phi
- \phi \left( \mathcal{D}_{\mu} \phi \right)^{*} \right] \nonumber \\
&=& \frac{q}{2} \left[ i \left( \phi^{*} \partial_{\mu} \phi
- \phi \partial_{\mu} \phi^{*} \right)
- 2 q A_{\mu} |\phi|^2 \right] \; . \hspace{5mm}
\end{eqnarray}

The energy--momentum tensor, which is the source of the gravitational field, includes contributions from both the complex scalar field and the electromagnetic fields, \mbox{$T_{\mu \nu} = \left(T_{\phi} \right)_{\mu \nu}+\left(T_{\mathcal{F}} \right)_{\mu \nu}$}, where the scalar and electromagnetic contributions are respectively:
\begin{eqnarray}
\label{TensorPhi}
\left( T_{\phi} \right)_{\mu \nu} &=& \frac{1}{2} \left[ \rule{0mm}{4mm}
\left( \mathcal{D}_{\mu} \phi \right)^* \left( \mathcal{D}_{\nu} \phi \right)
+ \left( \mathcal{D}_{\nu} \phi \right) \left( \mathcal{D}_{\mu} \phi \right)^* \right.
\nonumber \\
&-& \left. g_{\mu \nu} \left( \left( \mathcal{D}_{\lambda} \phi \right)
\left( \mathcal{D}^{\lambda} \phi \right)^* + m^{2} |\phi|^{2} \right) \right] \; ,
\\
\label{TensorFaraday}
\left(T_{\mathcal{F}} \right)_{\mu \nu} &=& \frac{1}{4\pi} \left(
\mathcal{F}_{\mu \lambda} {\mathcal{F}_\nu}^\lambda
- \frac{g_{\mu\nu}}{4} \: \mathcal{F_{\lambda \sigma}} \mathcal{F^{\lambda \sigma}}
\right) \; .
\end{eqnarray}

Variation of the action~\eqref{actionTOTAL} with respect to the different fields results in the Einstein field equations:
\begin{equation}
G_{\mu \nu} = 8 \pi T_{\mu \nu} \; ,    
\end{equation}
with $G_{\mu \nu}$ the Einstein curvature tensor of the space--time, plus the Mawxell equations:
\begin{equation}
\nabla_\mu \mathcal{F}^{\mu \nu} = - 4 \pi j^\nu \; , \qquad
\nabla_\mu \mathcal{F}^{* \mu \nu} = 0 \; ,
\label{Maxwell}
\end{equation}
where $\mathcal{F}^{* \mu \nu} := - \epsilon^{\mu \nu \alpha \beta} \mathcal{F}_{\alpha \beta}/2$ is the dual electromagnetic tensor (using the convention that $\epsilon^{0123}=-1/\sqrt{-g}$ and $\epsilon_{0123}=+\sqrt{-g}$), 
and the Klein--Gordon equation for the scalar field:
\begin{equation}
\left( \mathcal{D}^{\mu} \mathcal{D}_{\mu} - m^2 \right) \phi = 0 \; .
\label{KleinGordon}
\end{equation}
Notice that this is the usual Klein--Gordon equation except for the fact that it involves the gauge covariant derivatives defined above in Eq.~\eqref{DefsGaugeDev}.

\vspace{5mm}

A word about our conventions for the electromagnetic field is in order here. While quite common in the general relativity community, our conventions are by no means universal and many references, including the work of Jetzer and Van Der Bij in~\cite{Jetzer89}, define the electromagnetic Lagrangian with a factor $1/4$ instead of the $1/16 \pi$ used in the action~\eqref{actionTOTAL} above.  This implies that the definitions for the potential 1-form and the Faraday tensor, as well as the electric and magnetic fields that we will introduce below, have an extra factor of $1/(4 \pi)^{1/2}$ with respect to our convention, $\hat{A}_\mu = A_{\mu} / (4 \pi)^{1/2}$. In order to keep expressions like the gauge covariant derivative unchanged this requires one to also rescale the charge as $\hat{q} = (4 \pi)^{1/2} q$.  This implies, for example, that while with our convention the Newtonian force between two charged particles is simply $F=q_1 q_2/r^2$, with the alternative convention it would be $F=\hat{q}_1 \hat{q}_2/ (4 \pi r^2)$. Similarly, while with our convention a maximally charged Reissner--Nordstr\"om black hole would have a total charge equal to its total mass, $Q=M$, with the alternative convention this would correspond to $\hat{Q}/(4 \pi)^{1/2}=M$. In~\cite{Jetzer89} Jetzer and Van Der Bij also do not use the $1/2$ factor in the Lagrangian for the scalar field, so that their scalar field is rescaled with respect to ours as $\hat{\phi} = \phi / \sqrt{2}$ (they also use an opposite signature for the metric).

Finally, in order to simplify further their expressions,  Jetzer and Van Der Bij later rescale the charge as \mbox{$\tilde{q}=(\hat{q}/m)/[(8 \pi)^{1/2}] = (q/m)/\sqrt{2}$} (equation (14) in reference~\cite{Jetzer89}), which implies that while the critical charge for boson stars is expected to be $q_c=m$ with our convention, in their convention this corresponds to $\tilde{q}_c=1/\sqrt{2}$ (remember that we use Planck units so that $M_{PL}=1$). This convention has later been followed by other authors~\cite{Pugliese_2013,Kain_2021}, so we will use it below when we present our numerical results en section~\ref{sec:results}.


\section{Charged boson stars}
\label{sec:bosonstar}

In order to study stationary configurations corresponding to charged boson stars we assume spherical symmetry and the usual harmonic ansatz for the time dependence of the scalar field:
\begin{equation}
\label{ComplexFieldA}
\phi(t,r) = \phi_{0}(r)e^{i \omega t} \; ,
\end{equation}
with $\omega$ a real constant corresponding to the frequency of oscillation of the scalar field, and $\phi_0(r)$ a real-valued radial function which corresponds to the profile of the charged boson star~\cite{Liebling:2012fv,Jetzer89,Pugliese_2013}.  Both contributions to the energy--momentum tensor given by Eqs.~\eqref{TensorPhi} and \eqref{TensorFaraday} are independent of time under this harmonic ansatz. As a consequence, the space--time metric turns out to be static.  This implies that one can write the metric in spherical coordinates $(r,\theta,\varphi)$ in the polar-areal gauge as:
\begin{equation}
\label{METRIC}
ds^2 = - \alpha(r)^2 dt^2 + A(r) dr^2 + r^2 d\Omega^2 \; ,
\end{equation}
with $\alpha$ and $A$ functions only of the radial coordinate $r$,
and where \mbox{$d\Omega^2 = d\theta^2 + \sin^2(\theta) d\varphi^2$} is the standard solid angle element.

\vspace{5mm}

In the following we will consider the 3+1 formalism of general relativity, where the space--time $(\mathcal{M},g_{\mu\nu})$ is assumed to be globally hyperbolic, so that it can be foliated by a family of spacelike hypersurfaces $\Sigma_t$ that are parametrized by a global time function $t$, that is $\mathcal{M}\cong\mathbb{R}\times\Sigma_t$~\cite{Alcubierre08a,Gourgoulhon2012}. In this formalism the metric given by \eqref{METRIC} corresponds to taking a null shift vector $\beta^i=0$ and a spatial metric of the form $\gamma_{ij}=\mathrm{diag}(A,r^2,r^2 \sin(\theta))$, with $\alpha(r)$ the lapse function~\cite{Torres_2014}. With this form of the metric each spacelike hypersurface $\Sigma_t$ has a normal timelike unit vector field $n^\mu$ such that:
\begin{equation}
n^\mu = (1/\alpha,0,0,0) \; , \qquad
n_\mu = (-\alpha,0,0,0) \: .
\end{equation}
The timelike unit vector $n^\mu$ can be identified with the 4-velocity of observers moving along
the normal direction to the spacelike hypersurfaces, the so-called Eulerian observers.


\subsection{Gauss constraint}

Within the 3+1 formalism the covariant equations for the electromagnetic field can be formulated in terms
of the electric and magnetic fields as measured by the Eulerian observers (see e.g.~\cite{Alcubierre:2009ij}). In terms of the Faraday tensor $\mathcal{F}_{\mu\nu}$ and its dual, the electric $E^\mu$ and magnetic $B^\mu$ fields are defined as:
\begin{equation}
E^{\mu} = - n_{\nu} \mathcal{F}^{\mu \nu} \; , \qquad
B^{\mu} = - n_{\nu} \mathcal{F}^{* \mu \nu} \; .
\label{DefsElectricMagnetic}
\end{equation}

Since the electric and magnetic fields are purely spatial one can consider only their spatial components $E^i$ and magnetic $B^i$. Moreover, in spherical symmetry the electric field only has a non-zero component in the radial direction $E^i = (E , 0, 0)$, while the magnetic field vanishes identically $B^i=0$. This implies that the potential 1-form can be taken to be of the form $A_{\mu}=(A_0,0,0,0)$~\cite{Jetzer89,Kain_2021}.

Projecting the covariant Maxwell equations~\eqref{Maxwell} onto the normal vector $n^\mu$ we obtain two constraint equations to solve for the initial data~\cite{Alcubierre:2009ij}. However, the constraint equation for the magnetic field is trivial since this field vanishes. The only Maxwell equation to solve is therefore the Gauss constraint, which now takes the form~\cite{Torres_2014}:
\begin{equation}\label{MaxwellEqs2}
D_{i} E^i = 4 \pi e \; ,
\end{equation}
where $D_{i}$ is the covariant derivative compatible with the spatial metric $\gamma_{ij}$, and $e:=- n^\mu j_\mu$ is the electric charge density measured by the Eulerian observers:
\begin{eqnarray}
e &:=& - n^\mu j_\mu = - \frac{j_0}{\alpha} \nonumber \\
&=& - \frac{q}{2 \alpha} \left[ i \left(\phi^* \partial_t \phi
- \phi \partial_t \phi^* \right) - 2 q A_0 |\phi|^2 \right] \; .
\end{eqnarray}
Using now our harmonic ansatz for the scalar field, Eq. \eqref{ComplexFieldA}, the Gauss constraint can be written in terms of the metric functions and the electric field explicitly as:
\begin{equation}
\label{MaxwellEqsF}
\frac{dE}{dr} = 4 \pi q \phi_0^2 \left( \frac{\omega - qF}{\alpha} \right)
- \left( \frac{1}{2A} \frac{dA}{dr} + \frac{2}{r}  \right) E \; ,
\end{equation}
where we have defined the electric scalar potential as $F:= \alpha \Phi$, with $\Phi := - n^\mu A_\mu$. Since the potential 1-form has only a non-zero time component we immediately find $F=-A_{0}$. From this and Eq.~\eqref{DefsFaradayGaugeDev2} one also finds:
\begin{equation}
\frac{dF}{dr} = - \alpha A E \; .
\label{EqF}
\end{equation}
This is just the generalization of the usual expression for $E$ as minus the gradient of the electric potential to the case of our curved space--time.


\subsection{Klein--Gordon equation}

The Klein--Gordon equation~\eqref{KleinGordon} can be reduced to first order form by defining the variables:
\begin{equation}
\hat{\Pi} := n^{\mu} \mathcal{D}_\mu \phi^* \; , \qquad
\hat{\chi}_i := P^\mu_i \mathcal{D}_\mu \phi \; ,
\end{equation}
with $P^\mu_\nu = \delta^\mu_\nu + n^\mu n_\nu$ the projection operator onto the spatial hypersurfaces.
Using the metric~\eqref{METRIC} and the harmonic ansatz~\eqref{ComplexFieldA} one now finds:
\begin{equation}
\hat{\Pi} = i\left( \frac{\omega-qF}{\alpha} \right) \phi_0e^{i\omega t} \; , \qquad
\hat{\chi}_{i} = \left( \chi e^{i\omega t},0,0 \right) \; ,
\end{equation}
where now:
\begin{equation}
\chi = d\phi_0/dr \; .
\label{EqChi}
\end{equation}
From this, the Klein--Gordon equation can be rewritten as a first order differential equation for $\chi$:
\begin{equation}
\label{KleinGordonEqF}
\frac{d \chi}{dr} = - \chi \left( \frac{1}{\alpha} \frac{d \alpha}{d r}
+ \frac{1}{2A} \frac{dA}{d r} + \frac{2}{r} \right)
+ A \phi_{0} \left( m^2 - \left( \frac{\omega - qF}{\alpha} \right)^2 \right) \; .
\end{equation}


\subsection{Hamiltonian constraint and slicing condition}

Since the space--time is static the extrinsic curvature vanishes $K_{ij}=0$. This implies that the momentum constraints are trivial so we only need to solve the Hamiltonian constraint, which in this case reduces to:
\begin{equation}
^{(3)}R = 16 \pi \rho \; ,
\end{equation}
where $^{(3)}R$ is the Ricci scalar associated with the spatial metric $\gamma_{ij}$, and $\rho := n^{\mu} n^{\nu}T_{\mu \nu}$ is the local energy density measured by an Eulerian observers.  Using the energy--momentum tensor given by Eqs.~\eqref{TensorFaraday} and~\eqref{TensorPhi}, the Hamiltonian constraint takes the form of a first order differential equation for the radial metric function $A(r)$:
\begin{equation}
\frac{dA}{dr} = A \left\{ \rule{0mm}{6mm} \frac{1-A}{r} + r (AE)^2
+ 4 \pi r A \left[ \phi_0^2 \left( \frac{\omega - qF}{\alpha} \right)^2
+ \frac{\chi^2}{A} + \left( m \phi_0 \right)^2 \right] \right\} \; .
\label{HamEqF}
\end{equation}

\vspace{5mm}

We still need to find an equation for the lapse function $\alpha(r)$.  Since we are working in the polar-areal gauge~\cite{Seidel90,Bardeen83}, the natural slicing condition is precisely the so-called polar slicing
condition $\partial_t K_{\theta \theta}=K_{\theta \theta}=0$, which in this case reduces to:
\begin{equation}
\frac{d \alpha}{dr} = \alpha \left( \frac{A-1}{r} + \frac{1}{2A} \frac{dA}{dr}
- r (AE)^2  - 4 \pi r A (m \phi_0)^2 \rule{0mm}{5mm} \right) \; .
\label{LapseConditionF}
\end{equation}


\section{Total mass and charge}
\label{sec:mass-charge}

When working in terms of the areal radius it is well known that the total mass $M$ of the spacetime can simply be found as integral over a flat volume element of the form:
\begin{equation}
M = 4 \pi \int_0^\infty \rho r^2 dr \; ,
\label{TotalMass}
\end{equation}
where $\rho$ is the energy density of matter that we introduced before, $\rho = n^\mu n^\nu T_{\mu \nu}$. The above result can be shown directly from the Hamiltonian constraint. In our case, the energy density $\rho$ has contributions both from the scalar field and the electromagnetic field, and takes the form:
\begin{eqnarray}
\rho &=& \frac{1}{2} \left( \hat{\Pi}^* \hat{\Pi} + \frac{\chi_i^* \chi^i}{A} + m^2 \phi^2 \right)
+ \frac{A E^2}{8 \pi} \nonumber \\
&=& \frac{1}{2} \left[  \left( \frac{(\omega - q F)^2}{\alpha^2} + m^2 \right) \phi^2
+ \frac{\chi^2}{A} \right] + \frac{A E^2}{8 \pi} \; . \hspace{5mm}
\end{eqnarray}

The mass integral~\eqref{TotalMass} defined above, though correct, nevertheless has a serious drawback due to the fact that for a charged boson star (or indeed for any charged particle) the electric potential $F$ decays as $1/r$, so that the integral converges to the total mass $M$ rather slowly with $r$.  We will come back to this problem below.

The total electric charge $Q$ is provided by the conserved Noether charge which is defined by the local $U(1)$ symmetry. Integrating the time component of the conserved current $j^\mu$ we obtain:
\begin{equation}
Q = \int j^0 \sqrt{-\gamma} \: dx^{3} \; ,
\end{equation}
with $\gamma$ the determinant of the spatial metric. Substituting the value of $j^0$ using Eq.~\eqref{4Current}, and the harmonic ansatz, we obtain the following explicit expression for the charge integral:
\begin{equation}
Q = 4 \pi q \int_0^{\infty} \frac{\left(\omega - qF \right)}{\alpha}
\: \phi_0^{2} A^{1/2} r^2 dr \; .
\label{TotalCharge}
\end{equation}
In contrast to the total mass, the charge integral above converges very rapidly since for a boson star the scalar field $\phi_0$ decays exponentially (see following Section). It is also interesting to notice that while the mass integral~\eqref{TotalMass} involves a flat volume element, the charge integral~\eqref{TotalCharge} involves the full curved space volume element, hence the factor $A^{1/2}$ that appears in~\eqref{TotalCharge} but not in~\eqref{TotalMass}.

Since the total charge is related to the total number of particles $N$ as $Q = qN$, the same integration allows us to find $N$~\cite{Schunck:2003kk,Jetzer89}.  This is quite useful as one can use the total mass $M$ and the total number of particles $N$ to define a binding energy for the star as:
\begin{equation}
E_B := M - m N = M - (m/q) \: Q \; .
\label{BindingEnergy}
\end{equation}

In order to understand this definition notice first that the total mass $M$ includes all possible contributions to the energy of the star, that is it includes the rest-mass plus the kinetic and potential energies.  So, if we subtract from $M$ the total rest-mass given by $mN$, we are left with just the kinetic (positive) and potential (negative) contributions, which is precisely the definition of the binding energy.  If the binding energy is negative the system is gravitationally bound, while if it is positive the system is not bound, and even very small perturbations can cause it to dissipate to infinity.

\vspace{5mm}

One can also find the total mass of the system in an alternative way by assuming that
far away the metric reduces to the Reissner--Nordstr\"{o}m metric, so that:
\begin{equation}
A(r) \; \rightarrow \; \left(1-\frac{2M}{r} + \frac{Q^2}{r^2} \right)^{-1} \; .
\label{RelationINFTY1}
\end{equation}
Solving for $M$ we then find:
\begin{equation}
M = \lim_{r \rightarrow \infty} \; \left[ \frac{r}{2}
\left( 1 + \frac{Q^2}{r^2} - \frac{1}{A} \right) \right] \; .
\label{TotalMass2}
\end{equation}
Having first found the total charge $Q$ using~\eqref{TotalCharge}, we can use the above expression to find the total mass $M$. It turns out that this expression in fact converges very rapidly with $r$, since once we are in a region where the scalar field is negligible the space--time reduces to the pure electro-vac Reissner--Nordstr\"{o}m solution. Because of this we prefer to use~\eqref{TotalMass2} to obtain the total mass instead of the integral~\eqref{TotalMass}, but we have indeed checked that for our solutions both expressions agree for very large $r$.

As already mentioned, for a boson star the scalar field decays exponentially so that there is no real surface.  However, we can use the charge integral above to define an effective radius $R_{99}$ as that which contains $99\%$ of the total charge.  We could in principle do the same with the mass integral and define a (somewhat different) effective radius $R_{99}$ that contains $99\%$ of the total mass, indeed this is what is usually done for boson stars with no electric charge.  However, in light of the discussion above regarding the convergence rates of the mass and charge integrals, for charged boson stars it is in fact much better to define $R_{99}$ in terms of the charge integral, and this is what we will do when we report our results below.


\section{Boundary conditions, rescaling, and numerical methods}
\label{sec:boundaries}

The system of equations to be solved in order to construct charged boson star configurations consists on the Gauss constraint~\eqref{MaxwellEqsF}, the Klein--Gordon equation~\eqref{KleinGordonEqF}, the Hamiltonian constraint~\eqref{HamEqF}, and the polar slicing condition~\eqref{LapseConditionF}, together with the equation for the scalar potential~\eqref{EqF}, and the definition of $\chi$~\eqref{EqChi}, for the six functions $\{E,F,\phi_0,\chi,A,\alpha\}$.

To solve this system we must also choose appropriate boundary conditions to ensure that the solutions are regular at the origin and that the space--time is asymptotically flat. For the boundary conditions at the origin we take:
\begin{equation}
\label{BoundaryConditionsOrigin}
    \begin{aligned}
        & \alpha(0) = 1 \; , & \partial_r \alpha(0) = 0 \; , \\
        & A(0) = 1 \; ,      & \partial_r A(0) = 0 \; , \\
        & \phi_0(0) = k \; , & \partial_r \phi_0(0) = 0 \; , \\
        & F(0) = 0 \; ,      & \partial_r F(0)  = 0 \; ,
    \end{aligned}
\end{equation}
with $k>0$ a positive real constant. The vanishing of the radial derivatives at the origin is due to the spherical symmetry.  Asking for $A(0)=1$ is required in order to guarantee that the space--time is locally flat there, while the constant value of $\phi_0(0)=k$ is our free parameter. On the other hand, choosing $\alpha(0)=1$ and $F(0)=0$ is done just for simplicity, since we don't know the correct values of those variables there (though one could argue that there are no ``correct'' values there since these are just gauge functions).  But we will have more to comment on these choices below. Notice in particular that with these conditions we also have $E(0)=\chi(0)=0$.

For solutions that represent an isolated star the scalar field must also vanish at infinity, that is $\phi_0(r)\rightarrow 0$ for $r\rightarrow \infty$. For each choice of $\phi_0(0)$, our system of equations has solutions that decay exponentially at infinity only for certain frequencies $\omega$. This means that, given a central value of the scalar field, we must solve an eigenvalue problem in order to find the frequency $\omega$.  Notice that the Klein--Gordon equation~\eqref{KleinGordonEqF} implies, in particular, that for large $r$ the following condition must be satisfied in order to have exponentially decaying solutions for the scalar field:
\begin{equation}
m \geq (\omega - q F_\infty)/\alpha_\infty \; ,
\label{omega-condition}
\end{equation}
with $\alpha_\infty$ and $F_\infty$ the asymptotic values of $\alpha$ and $F$. If this condition is not satisfied one would have instead sinusoidal solutions for the scalar field for large $r$, that are simply not compatible with an asymptotically flat space--time.

Given a value of $\phi_0(0)=k$ as a free parameter, we choose a trial value of the frequency $\omega$ and integrate our system of equations outwards from the origin using standard fourth order Runge--Kutta. We then use a shooting algorithm to find the correct value of $\omega$ that corresponds to exponential decay of the scalar field far away (the value of $\omega$ found in this way is typically such that $\omega>m$, but this changes once we apply the rescaling described below). We also look for solutions with no nodes in the scalar field, corresponding to the ground state of our charged boson stars. One can also solve for excited states with one or mode nodes, but we will discuss such solutions elsewhere.

\vspace{5mm}

Going back to our boundary condition for $\alpha$ at the origin, we now notice that in the final solution we in fact do not want \mbox{$\alpha(0)=1$}, but rather $\alpha(r) \rightarrow 1$ at infinity, corresponding to Minkowski space--time. But this is no problem as the polar slicing condition~\eqref{LapseConditionF} is linear in $\alpha$, so we can always just rescale the lapse. However, in order not to affect the solution, we must also rescale the frequency $\omega$ and the scalar electric potential $F$ by the same factor, as the whole system of equations is easily shown to be invariant under the change:
\begin{equation}
\alpha \rightarrow \alpha/C_1 \; , \qquad
\omega \rightarrow \omega/C_1 \; , \qquad
F \rightarrow F/C_1 \; ,
\end{equation}
with $C_1$ an arbitrary constant.  We choose the constant $C_1$ by extrapolating the value of the lapse $\alpha$ at infinity assuming an asymptotic behavior of the form \mbox{$\alpha \sim \alpha_\infty + cte/r$}, so that $C_1=\alpha_\infty$. This ensures that after rescaling the lapse will now go to 1 at infinity.

For a non-charged boson star this rescaling gives us the final ``physical'' frequency, but for the charged case we are not yet done.  Our solution was also found by asking for the electric potential to satisfy $F(0)=0$, and this remains true even after the rescaling above.  But it would seem much more natural to ask instead for $F(r) \rightarrow 0$ at infinity. We can fix this by making a gauge transformation as in Eq.~\eqref{GaugeTrans}, with an appropriately chosen gauge function of the form $\theta = C_2 t$, with $C_2$ constant.  The gauge transformation in this case simplifies to:
\begin{equation}
\phi \rightarrow \phi \: e^{i q C_2 t} \; , \qquad
F \rightarrow F + C_2 \; .
\end{equation}
As before, we now choose the constant $C_2$ by extrapolating the value of $F$ at infinity assuming an asymptotic behavior of the form $F \sim F_\infty + cte/r$, so that $C_2=-F_\infty$. The gauge
transformation above clearly implies that the frequency must also be transformed as:
\begin{equation}
\omega \rightarrow \omega + q C_2 \; .
\end{equation}
After these two transformations our final solution is now such that $\alpha(r)\rightarrow 1$ and $F(r)\rightarrow 0$ at infinity, as desired.

For a given value of $\phi_0(0)=k$ we then have three different values for the frequency:  an initial value $\omega_1$ obtained by the shooting algorithm using our original boundary conditions $\alpha(0)=1$ and $F(0)=0$; a second value $\omega_2$ obtained after rescaling the lapse; and a final ``physical'' value $\omega_3$ obtained after performing the gauge transformation. In all our figures and tables below we always report this final value for the frequency. Notice also that since in our final solution the lapse function $\alpha$ goes to 1 at infinity and the electric potential $F$ goes to 0, condition~\eqref{omega-condition} reduces simply to $\omega \leq m$. We find that indeed this is always the case for all our solutions.


\section{Numerical results}
\label{sec:results} 

In this section we will present our results for the different families of charged boson stars. For simplicity, in all our solutions we have set the mass parameter to $m=1$, but the results can be easily rescaled to arbitrary values of $m$ since the system of equations is invariant under the transformation:
\begin{equation}
\begin{aligned}
m \rightarrow \lambda m \; , \quad
q \rightarrow \lambda q \; , \quad
\omega \rightarrow \lambda \omega \; , \\
r \rightarrow r/\lambda \; , \quad
\chi \rightarrow \lambda \chi \; , \quad
E \rightarrow \lambda E \; ,
\end{aligned}
\end{equation}
with $(\alpha,A,\phi_0,F)$ unchanged. The crucial parameter, however, is the charge to mass ratio $q/m$ which remains invariant under this transformation. Each configuration for a charged boson star can then be characterized by the central value of the scalar field $\phi_0(0)$ and its charge parameter $q$. In this paper we are also only considering the ground state, that is, we look for solutions with no nodes on the scalar field. These correspond to those solutions with the lowest possible value of the frequency $\omega$ for a given boson charge $q$ and central field amplitude $\phi_0(0)$.

As mentioned in the introduction, all the equations presented in the previous sections use a normalization such that the critical mass should be $q_c=m$.  However, in order to make our results easier to compare with previous works~\cite{Jetzer89,Pugliese_2013,Kain_2021}, in this section we will renormalize the charge as $\tilde{q} = (q/m) / \sqrt{2}$. Note that with this normalization the critical charge is now simply $\tilde{q}_c=1/\sqrt{2}$.


\subsection{Families of solutions for different values of the charge}

\begin{figure}[t]
\includegraphics[width=0.8\textwidth]{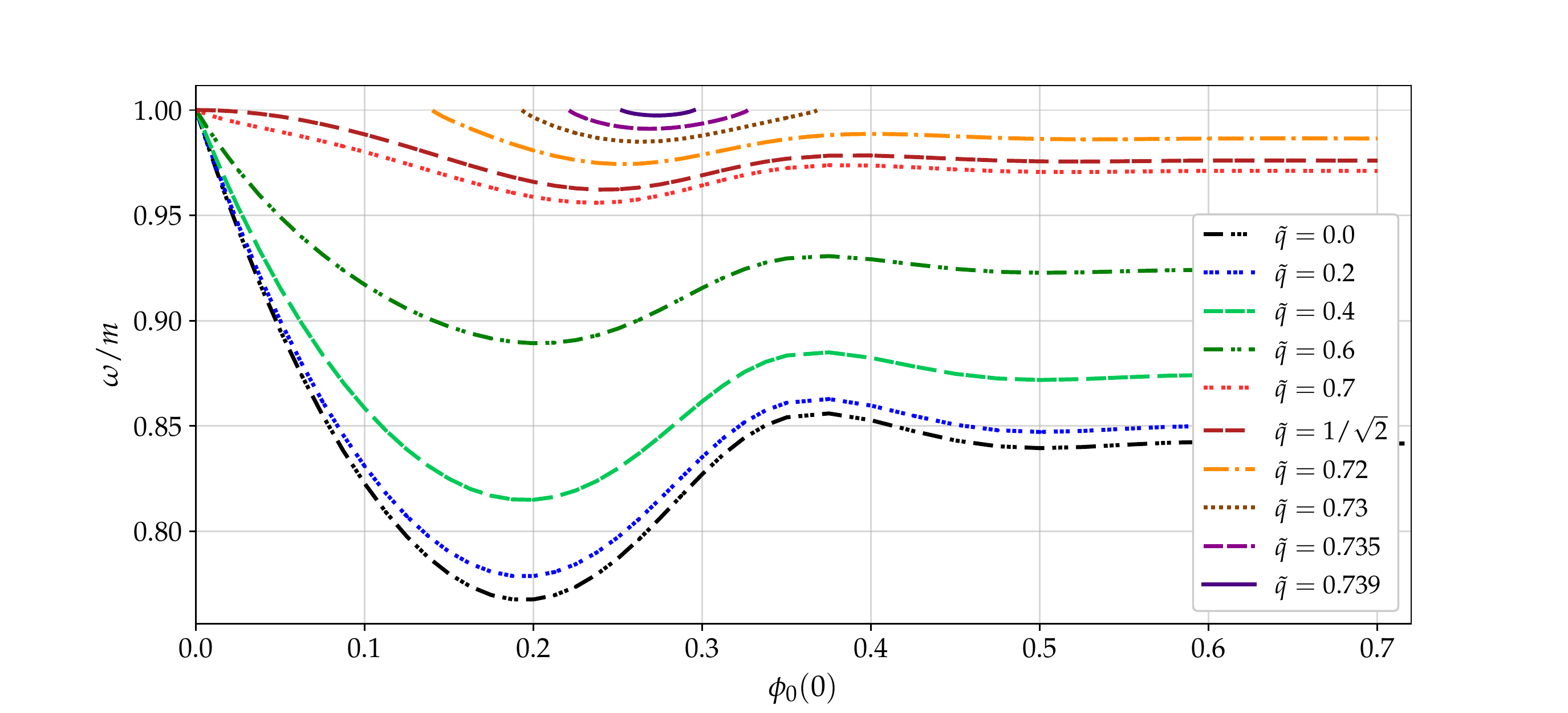}
\caption{Frequency $\omega$ for the charged boson star solutions as a function of the central scalar field value $\phi_0(0)$. Different color curves represent families of solutions for different values of the charge $\tilde{q}$. As $\tilde{q}$ increases, the frequencies $\omega$ increase to values closer to the value for the mass $m$. For super-critical charges $\tilde{q}>1/\sqrt{2}$ we are still able to find solutions, but only for a limited range of values of $\phi_0(0)$. This allowed range becomes narrower as we increase the charge further, until it disappears for $\tilde{q} \gtrsim 0.739$.}
\label{fig:phi_omega_q}
\end{figure}

We will now show our main results for the families of solutions corresponding to different values of the charge parameter $\tilde{q}$. Figure~\ref{fig:phi_omega_q} shows the relation between the frequency $\omega$ and the central value of the scalar field $\phi_0(0)$, for different values of $\tilde{q}$. We show families of solutions that  cover a wide range of values of the charge, from $\tilde{q}=0$ corresponding to the standard (mini) boson stars, all the way up to the critical charge $\tilde{q}=1/\sqrt{2}$, and even slightly above this value. For values of the charge such that $\tilde{q} \leq 1/\sqrt{2}$ it is in fact possible to solve the EMKG system for all values of $\phi_0(0)$. Notice, however, that for sub-critical charges the derivative of the frequency $\omega$ is always negative as we approach $\phi_0(0)=0$, that is $\left. (d\omega/d\phi_0) \right|_{\phi_0(0)=0}<0$ , while for the critical charge $\tilde{q}=1/\sqrt{2}$ we find instead $\left. (d\omega/d\phi_0) \right|_{\phi_0(0)=0} \simeq 0$, so that a local maximum seems to be developing at $\phi_0(0)=0$.

Much more interesting is the fact that for super-critical charges $\tilde{q}>1/\sqrt{2}$ we are still able to find solutions, but only for a limited range of values of $\phi_0(0)$, and a limited range of $\tilde{q}$ such that $1/\sqrt{2} \simeq 0.707 < \tilde{q} \lesssim 0.739$. This finite range is determined by the fact that no exponentially decaying solutions can exist for $\omega>m$. It also turns out that the allowed range is always bounded away from $\phi_0(0)=0$, that is there are no super-critical solutions with very small amplitudes.  As we increase the charge the allowed range of values for $\phi_0(0)$ becomes smaller and smaller, until for $\tilde{q} \gtrsim 0.739$ it disappears completely and no more solutions are found.

We should emphasize here that when finding solutions one must really make sure that they decay exponentially far away by moving the numerical integration boundary to large radius.  Is is quite easy to ``find'' what seems to be a nice decaying solution with the boundary relatively close by, only to discover that when we move the boundary further out the apparent exponential decay turns into a sinusoidal oscillation with small amplitude and large wavelength.  Reference~\cite{Pugliese_2013}, for example, reports finding supercritical solutions with charge as high as $\tilde{q}=0.8$, which according to our results would require $\omega>m$, and which we simply have been unable to reproduce.  We believe that such apparent solutions are no real solutions at all, and would have an asymptotic sinusoidal behavior if they were to be extended to large radius. The solutions in~\cite{Pugliese_2013} are also shown only up to a radius $r \sim 10$, which is still quite small for many of the configurations studied here.  This is confounded by the fact that they seem to have rescaled their frequencies using the behavior of the lapse and scalar potential at a finite radius (presumably the boundary of their computational domain) instead of their asymptotic behavior at infinity, which makes it very difficult to reproduce the frequencies they report.

\begin{figure}[t]
\centering
\includegraphics[width=0.8\textwidth]{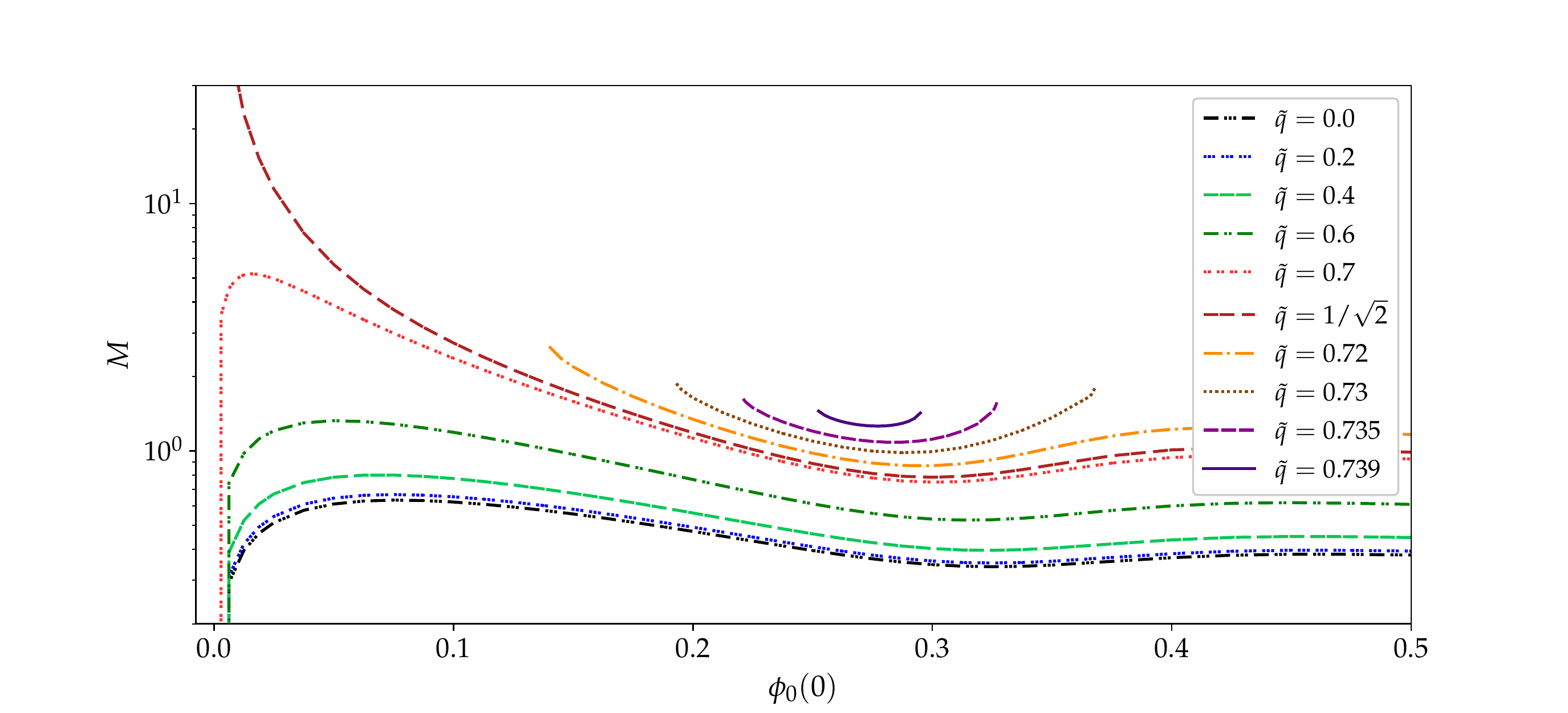}
\caption{Total mass (in units of $M_{PL}^2/m$) as a function of the central value of the field $\phi_0(0)$ for families of boson star solutions with different values of the charge $\tilde{q}$ (notice that the vertical axis is on a logarithmic scale).}
\label{fig:phi__M}
\end{figure}

Figure~\ref{fig:phi__M} shows the total mass $M$ (in units of $M_{PL}^2/m$) for the different families of solutions as a function of the central value of the scalar field $\phi_0(0)$. For boson stars with no charge, $\tilde{q}=0$, it is well known since the work of Kaup in 1968~\cite{Kaup68}, and Ruffini and Bonazzolla in 1969~\cite{Ruffini69}, that the mass has a maximum value $M_{\mathrm{Kaup}} \approx 0.633 \: M_{\mathrm{PL}}^2/m$. For central values of the field to the right of this maximum the total mass oscillates slightly and then converges to a value $M \sim 0.37 \: M_{\mathrm{PL}}^2/m$ for very large $\phi_0(0)$. From the Figure it is to easy see that this behavior of the mass is in fact very similar for charged boson stars as long as $\tilde{q}<1/\sqrt{2}$. The main change is that the maximum mass increases with $\tilde{q}$, while its position moves to lower values of $\phi_0(0)$ (this was also found by Jetzer and Van Der Bij in~\cite{Jetzer89}). Perhaps more interesting is the fact that the position of the maximum moves to $\phi_0(0) \rightarrow 0$ as the charge approaches the critical charge $\tilde{q} \rightarrow 1/\sqrt{2}$, while the value of the maximum mass diverges $M_{\mathrm{max}} \rightarrow \infty$.  This might seem surprising at first, but we must remember that as the central value of the field approaches zero the effective radius of the boson stars increases without bound, so even if the energy density decreases the total mass can still grow. As mentioned before, we also find solutions for slightly super-critical charges in a narrow range of $\phi_0(0)$.  For such super-critical solutions the mass has a local minimum within this allowed interval, and no local maximum is found.

For boson stars with no electric charge, $\tilde{q}=0$, the central value of $\phi_0(0)$ for which the mass reaches its maximum is known to separate stable configurations (to the left) from unstable ones (to the right). One would expect a similar thing to happen for charged boson stars.  This suggests that all our super-critical solutions, with no maximum and only a local minimum, should be unstable. Of course, in order to be sure of this one would need to do either a linear stability analysis or a full non-linear dynamical evolution (we will consider the dynamical simulation of charged boson stars in a future work).

\begin{figure}
\centering
\includegraphics[width=0.8\textwidth]{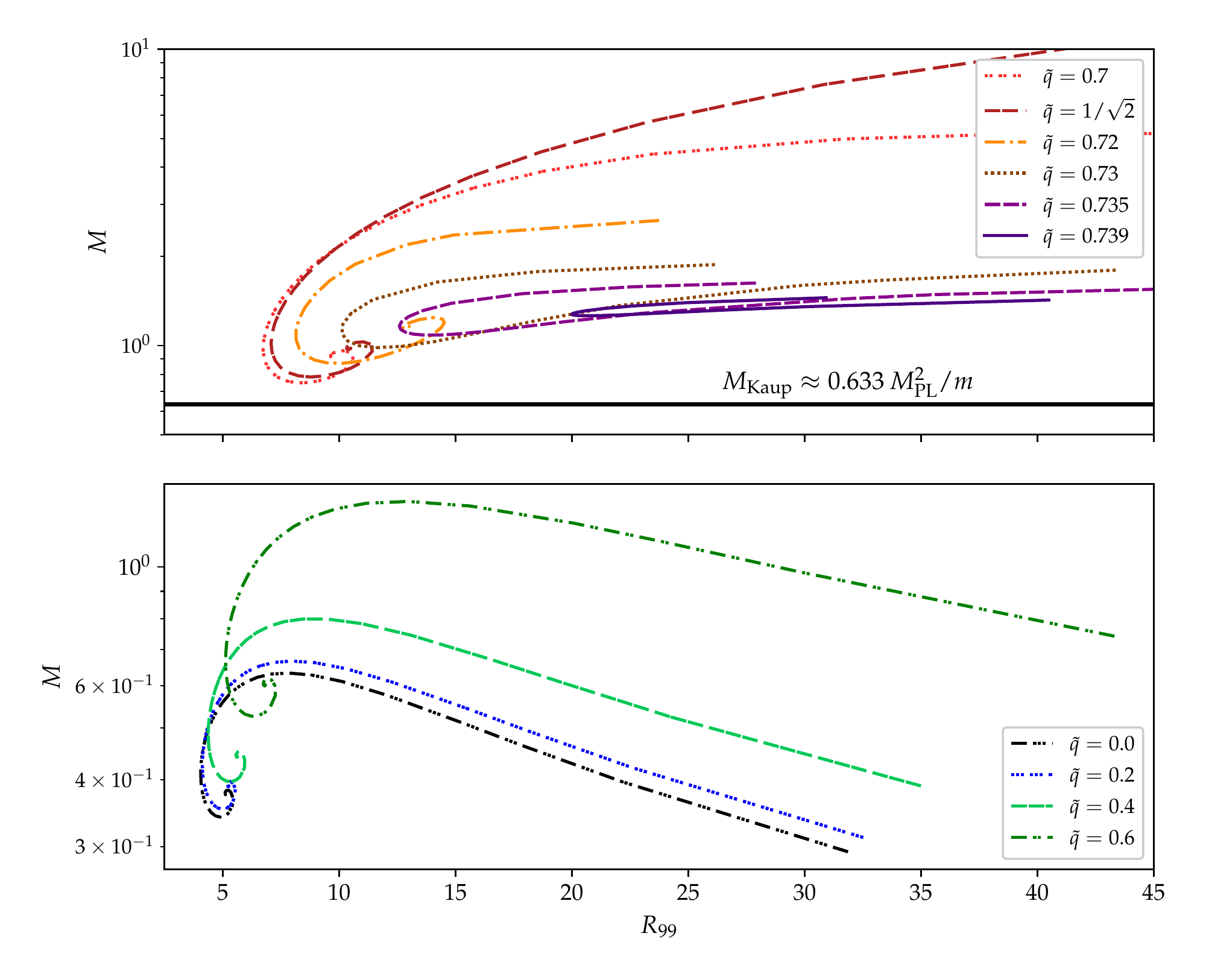}
\caption{ Total mass (in units of $M_{PL}^2/m$) versus effective radius $R_{99}$ for the different families of boson stars (notice that the vertical axis is on a logarithmic scale). {\em Lower panel}: sub-critical charge.  {\em Upper panel}: super-critical charge. We have included the sub-critical case $\tilde{q}=0.7$ in the upper panel in order to show that for $\tilde{q} \gtrsim 0.7$ the total mass $M$ for all configurations is always larger than the maximum for a $\tilde{q}=0$ boson star $M_\mathrm{Kaup}$.}
\label{fig:R_99__M}
\end{figure}

In Figure~\ref{fig:R_99__M} we show the relation between the total mass $M$ and the effective radius $R_{99}$. We have divided this Figure into two separate plots in order to better appreciate the changes in behavior close to and above the critical charge. As before, from the Figure it is clear that although the mass and effective radius increase with $\tilde{q}$, the behavior is similar for charges such that $0 \leq \tilde{q} < 1/\sqrt{2}$ (lower panel). On the other hand, for super-critical charges $1/\sqrt{2} \leq \tilde{q} < 0.739$ the behavior changes and there in no longer a local maximum for the mass (upper panel). The system with $\tilde{q}=0.7$ was included in the upper panel in order to show that for $\tilde{q} \gtrsim 0.7$ the total mass $M$ for {\em all}\/ configurations is always larger than the maximum mass for a $\tilde{q}=0$ boson star $M_\mathrm{Kaup}$.

\begin{figure}
\centering
\includegraphics[width=0.8\textwidth]{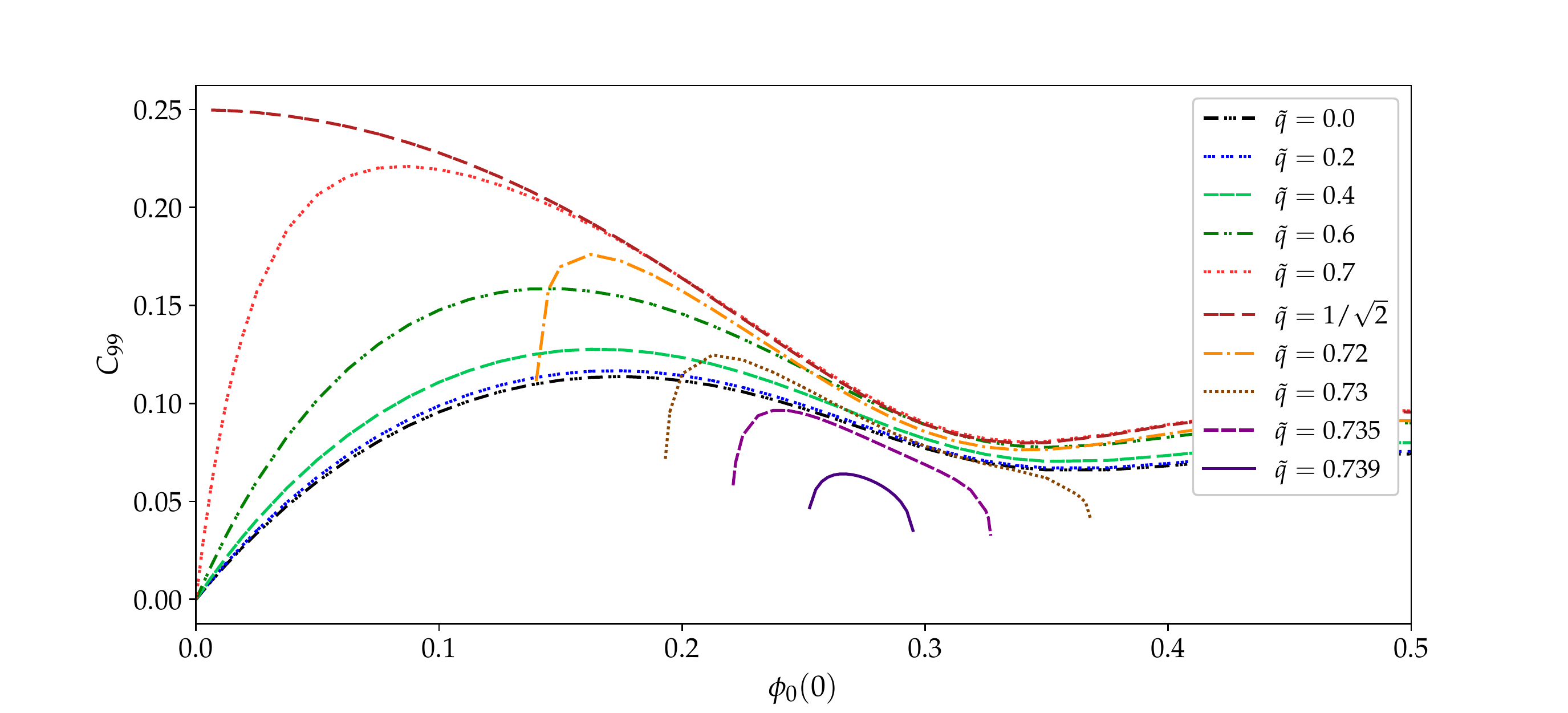}
\caption{Compactness $C_{99}:=M/R_{99}$ as a function of the central value of the scalar field $\phi_0(0)$. The compactness for boson stars with critical charge behaves as an upper bound for all other configurations.}
\label{fig:phi_C}
\end{figure}

Figure~\ref{fig:phi_C} shows the compactness of the boson stars defined as $C_{99}:=M/R_{99}$, as a function of the central value of the scalar field $\phi_0(0)$. Again, for sub-critical charges the behavior is very similar to that for standard $\tilde{q}=0$ boson stars. Notice, however, that for the family with critical charge $\tilde{q}=1/\sqrt{2}$ the compactness reaches a maximum of $C_{99} \rightarrow 0.25$ as $\phi_0(0) \rightarrow 0$ (compare this with the compactness for a Schwarzschild black hole $C=1/2$, and for a maximally charged Reissner-Nordstr\"om black hole $C=1$).  This might seem counter intuitive as in the case $\phi_0(0)=0$ there is no boson star, but remember that as we approach that limit the mass and effective radius both diverge.  This maximum compactness turns out to be an upper limit for all configurations. Notice also that for super-critical solutions with $\tilde{q}>1/\sqrt{2}$ the maximum compactness falls again.

\begin{figure}
\centering
\includegraphics[width=0.8\textwidth]{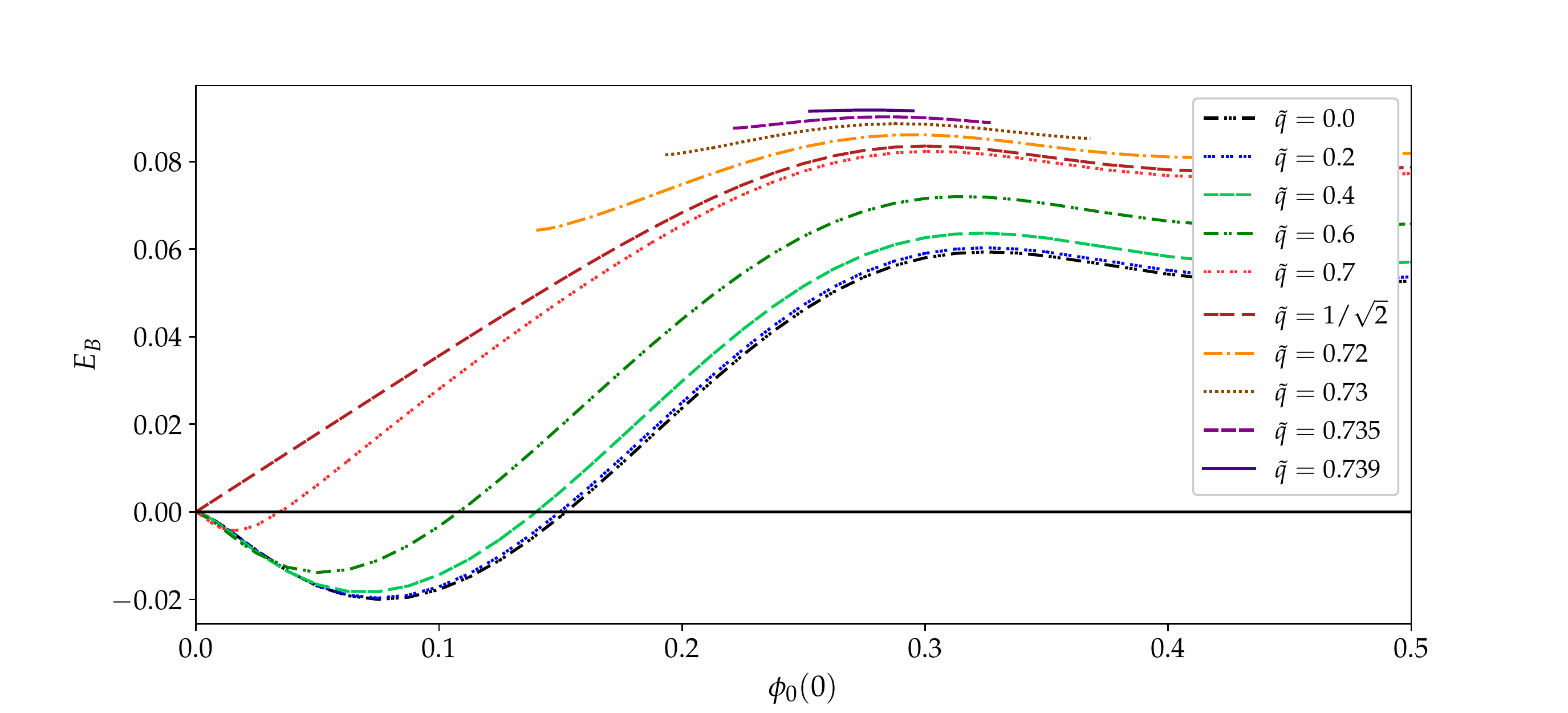}
\caption{Binding energy (in units of $M_{PL}^2/m$) for the charged boson stars configurations as a function of the central value of the scalar field $\phi_0(0)$. We can see that all sub-critical families have a region with negative binding energy, while for configurations with $\tilde{q} \geq 1/\sqrt{2}$ this region disappears.}
\label{fig:phi__E_B}
\end{figure}

In Figure~\ref{fig:phi__E_B} we show the binding energy $E_{B}:=M-mN$ for the different families of solutions, with $M$ the total mass and $N$ the total number of bosons.  Solutions with negative binding energy are gravitationally bound, while those with positive binding energy are unbound. Notice that gravitationally bound solutions can be either stable or unstable depending on whether the value of $\phi_0(0)$ is to the left or the right of that which corresponds to the maximum mass for that family.  On the other hand, unbound solutions with $E_B>0$ are all expected to be unstable. From the Figure we can see that all sub-critical families have a region with negative binding energy, though this region becomes smaller and smaller as we approach the critical charge.  However, for all solutions with $\tilde{q} \geq 1/\sqrt{2}$ the binding energy is always positive indicating that all such solutions are gravitationally unbound and therefore almost certainly unstable. From dynamical simulations of standard $\tilde{q}=0$ boson stars we know that unstable but bound configurations can either collapse to a black hole or migrate to a stable solution when perturbed, whereas unstable and unbound configurations either collapse to a black hole or disperse (explode) away to infinity. We expect a similar behavior for charged boson stars, but this will be studied in detail in a future work.

\begin{figure}[t]
\centering
\includegraphics[width=0.8\textwidth]{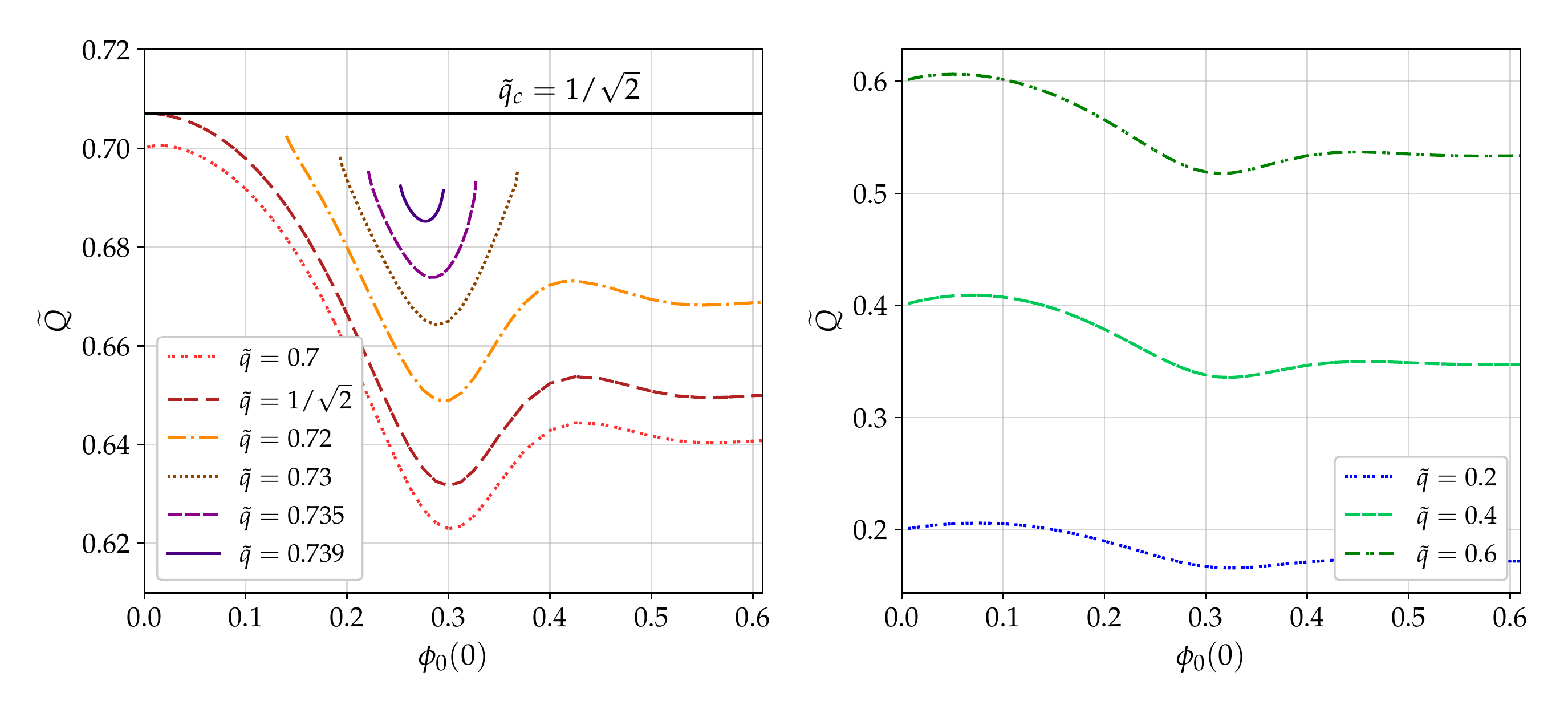}
\caption{ Renormalized total charge $\tilde{Q}:=(Q/M)/\sqrt{2}$ as a function of the central field amplitude $\phi_0(0)$. {\em Right panel}: Sub-critical cases with $\tilde{q}<1/\sqrt{2}$. {\em Left panel}: Super-critical cases with $\tilde{q} \geq 1/\sqrt{2}$ . We have included the sub-critical case $\tilde{q}=0.7$ in the left panel to show that the critical charge determines a upper bound of $\tilde{Q}/M$ for all solutions. }
\label{fig:QM__phi}
\end{figure}

There is another interesting relation one can find.  Notice that the total number of bosons $N$ can be written in terms of the total charge Q as $N=Q/q$, so that the binding energy becomes $E_B=M[1-(m/q)(Q/M)]$. This means that for gravitationally bound solutions we must have  $Q/M > q/m$, while for the unbound solutions we have $Q/M < q/m$. For consistency, we have also renormalized the total charge with a factor of $\sqrt{2} M$, so that we now define $\tilde{Q}:=(Q/M)/\sqrt{2}$. Notice that $\tilde{q}$ has already been defined as $\tilde{q}=(q/m)/\sqrt{2}$, so that comparing $Q/M$ with $q/m$ is now equivalent to comparing $\tilde{Q}$ directly with $\tilde{q}$.

Figure~\ref{fig:QM__phi} shows the renormalized charge $\tilde{Q}$ as a function of the central value of the scalar field $\phi_0(0)$.  We have divided Figure~\ref{fig:QM__phi} into two separate plots. In the right panel corresponds to families of boson stars with a sub-critical charge $\tilde{q}<1/\sqrt{2}$. The behavior is similar for all cases: there is region with $\tilde{Q} > \tilde{q}$ to the left of the plot, and another with $\tilde{Q} < \tilde{q}$ to the right. These regions correspond to the gravitationally bound and unbounded solutions respectively, as mentioned above. On the other hand, the left panel of the Figure shows families of boson stars with a super-critical charge $\tilde{q} > 1/\sqrt{2}$, including the sub-critical case $\tilde{q} = 0.7$. As we can expect, the behavior for the case with $\tilde{q} = 0.7$ is still the same as for the other sub-critical cases on the right panel, but the region  with $\tilde{Q} > \tilde{q}$ corresponding to gravitationally bound solutions is now almost negligible.  For the critical case $\tilde{q}=1/\sqrt{2}$ the region with $\tilde{Q} > \tilde{q}$ has now vanished completely and all solutions are unbound. Furthermore, we find that the limit determined by the critical charge $\tilde{q}_{c} = 1/\sqrt{2}$ (black line) acts as an upper bound for all solutions, that is we find that $\tilde{Q} \leq 1/\sqrt{2}$ for all the families of charged boson stars. From the definition of $\tilde{Q}$ we can in fact see that this upper bound corresponds to having precisely $Q=M$. This means that even though we have found super-critical solutions in the sense that $q>m$ ($\tilde{q}>1/\sqrt{2}$), there are {\em no}\/ super-critical solutions in the sense that $Q>M$ ($\tilde{Q}>1/\sqrt{2}$), which is precisely what one would have expected on physical grounds. That is, in general relativity it is not so much the ratio between the charge and mass parameters $q$ and $m$ the one that determines the existence of solutions, but rather the ratio between the total charge $Q$ and the total mass $M$.


\subsection{Some particular configurations}

\begin{figure}
\centering
\includegraphics[width=0.8\textwidth]{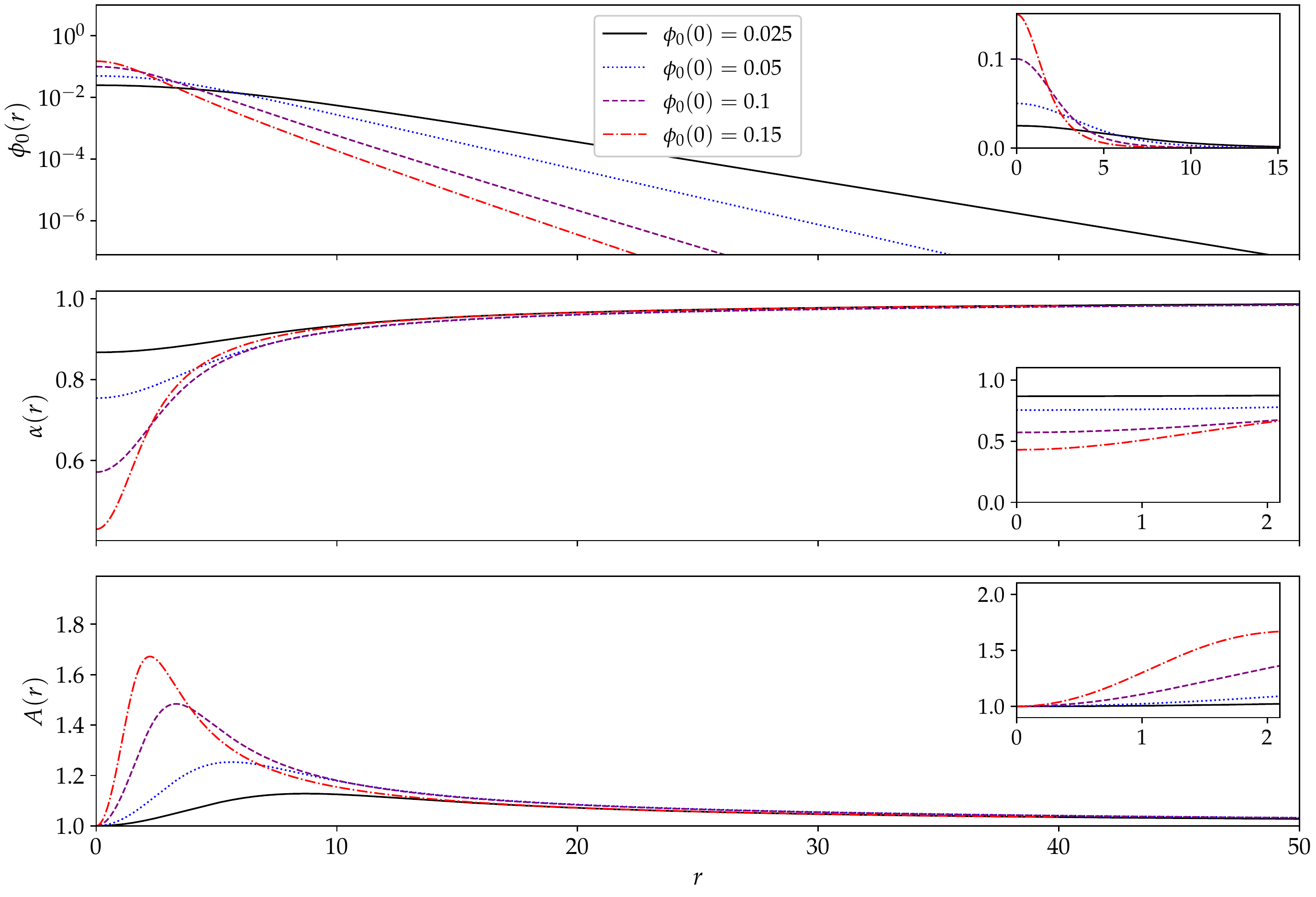}
\caption{Charged boson stars solutions corresponding to configurations with charge $\tilde{q}=0.4$ and different central amplitudes $\phi_0(0)=0.025,0.05,0.1,0.15$. The top, middle and lower panels show respectively the scalar field $\phi_0(r)$ (in a logarithmic scale), the lapse function $\alpha$, and the radial metric $A(r)$. The insets show a closeup or the behavior of the different functions close to the origin.}
\label{fig:field_0_4}
\end{figure}

We will now show examples of some particular configurations for charged boson stars in order to better understand the behavior of the different functions in terms of the radial coordinate $r$. Figure~\ref{fig:field_0_4} shows solutions for four boson stars with charge $\tilde{q}=0.4$, and different central values of the scalar field $\phi_0(0)=0.025,0.05,0.1,0.15$. The top panel shows the scalar field $\phi_0(r)$ as a function of the radial coordinate $r$. Notice that the scalar field is plotted in a logarithmic scale in  order to show that the field does decay exponentially far away, which in a logarithmic scale corresponds to a straight line. The shape of the field does not change significantly for the different central values $\phi_0(0)$, but the solution becomes more compact and cuspy for larger values of $\phi_0(0)$. Notice also that for $\phi_0(0)=0.025$ the exponential decay only becomes apparent for very large values of $r$, as already mentioned above. The middle panel of the Figure shows the lapse function $\alpha(r)$, and the lower panel shows the radial metric function $A(r)$. From the Figure we can clearly see that the lapse and radial metric satisfy the boundary conditions at infinity, $\alpha(r) \rightarrow 1$ and $A(r) \rightarrow 1$, and also $A(r=0)=1$. Notice that as $\phi_0(0)$ increases the lapse becomes smaller at the origin, while the radial function has a larger and larger maximum that also moves closer to the origin.  

\begin{figure}
\centering
\includegraphics[width=0.8\textwidth]{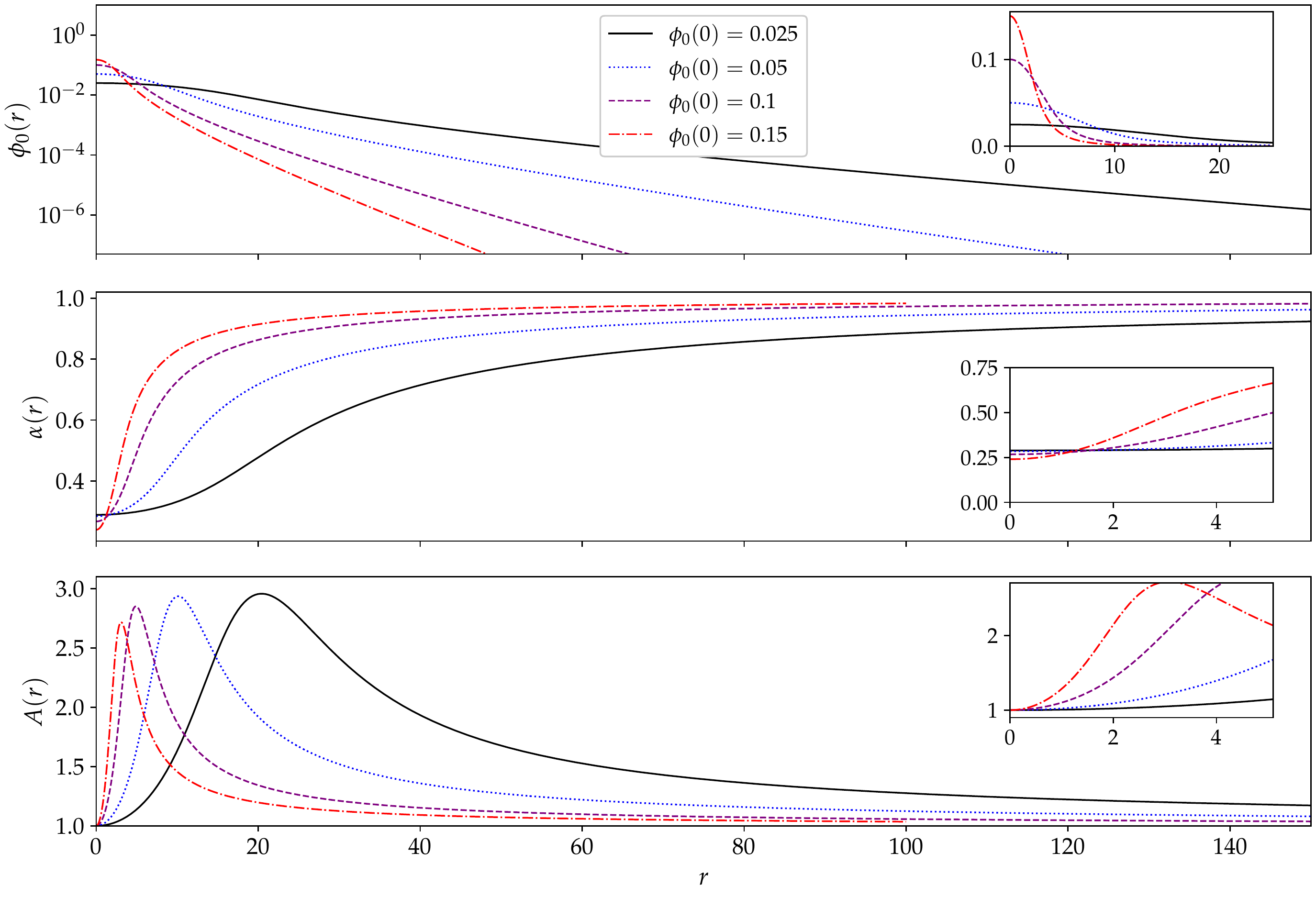}
\caption{Charged boson star solutions corresponding to configurations with the critical charge $\tilde{q}=1/\sqrt{2}$, and different central amplitudes $\phi_0(0)=0.025,0.05,0.1,0.15$.  The top, middle and lower panels show respectively the scalar field $\phi_0(r)$ (in a logarithmic scale), the lapse function $\alpha$, and the radial metric $A(r)$.}
\label{fig:field_0_crit}
\end{figure}

Figure~\ref{fig:field_0_crit} shows solutions for configurations with the critical charge $\tilde{q}=1/\sqrt{2}$, and different central values of the scalar field $\phi_0(0)=0.025,0.05,0.1,0.15$. Although the overall shape is very similar to that of the previous Figure, we can notice that the scalar field distribution is in general wider than in the solutions with $\tilde{q}=0.4$, i.e. the effective radius is larger (notice the change of scale on the horizontal axis). This implies that for smaller values of the central amplitude we in fact need to integrate to much larger values of $r$ in order to see the correct exponential decay. The increase in the width of the solution is most probably due to the electromagnetic repulsion. 

\begin{figure}
\centering
\includegraphics[width=0.8\textwidth]{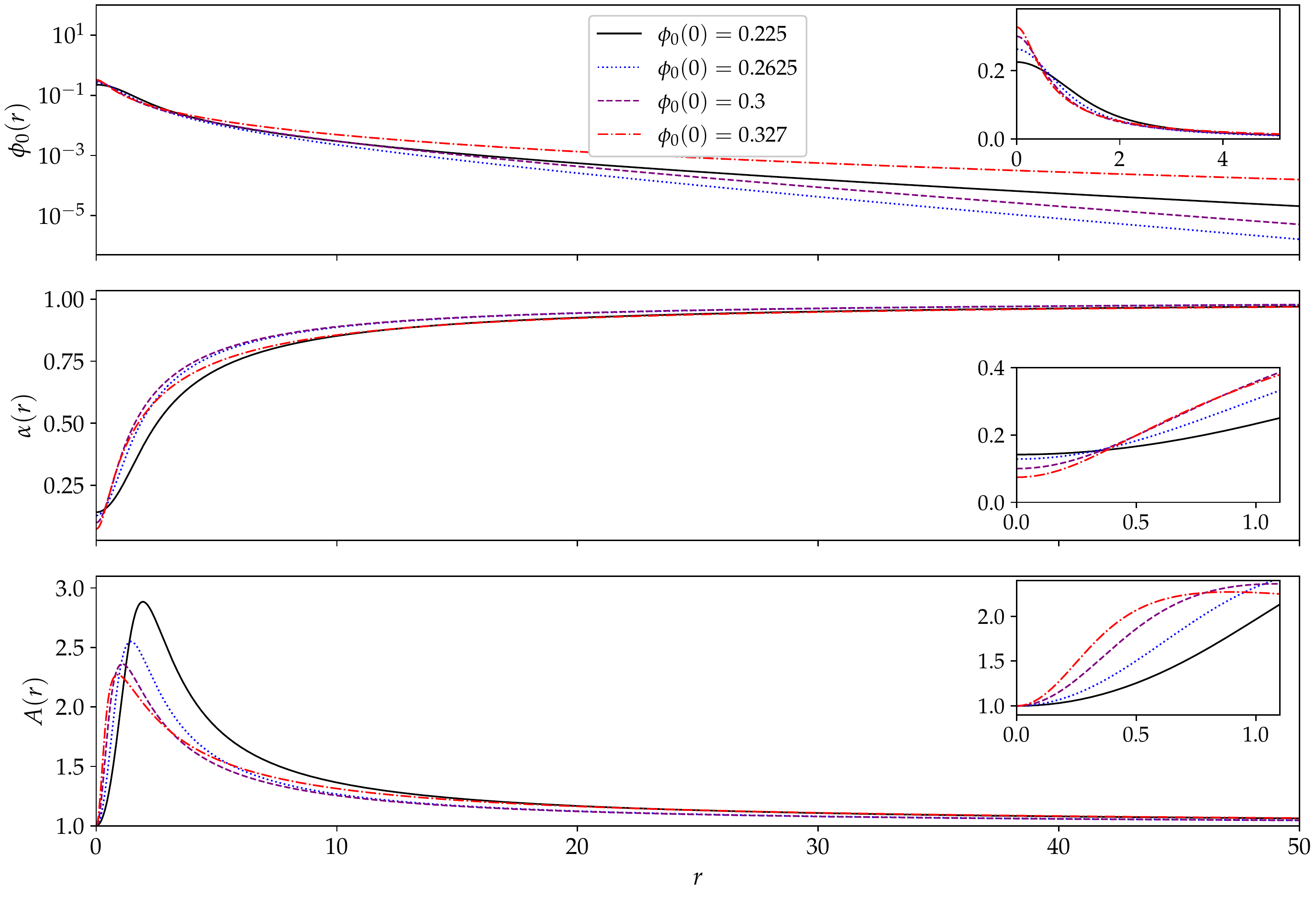}
\caption{Charged boson stars solutions corresponding to configurations with a super-crticial charge $\tilde{q}=0.735$, and different central amplitudes $\phi_0(0)=0.225,0.2625,0.3,0.327$. The top, middle and lower panels show respectively the scalar field $\phi_0(r)$ (in a logarithmic scale), the lapse function $\alpha$, and the radial metric $A(r)$.}
\label{fig:field_0_735}
\end{figure}

As our last example, in Figure~\ref{fig:field_0_735} we show solutions for configurations with a super-critical charge $\tilde{q}=0.735$. The behavior of the different functions is again similar to the previous cases, though we can see that the scalar field distribution is now even wider. From our previous discussion we know that super-critical solutions are only allowed for a small range of values of the central amplitude, which in this case corresponds to $0.22 \lesssim \phi_0(0) \lesssim 0.328$. The plot then shows the cases with $\phi_0(0)=0.225,0.2625,0.3,0.327$.  We can see that in all these cases the scalar field does indeed decay exponentially for large $r$.


\subsection{Time evolutions}

\begin{figure}[ht]
\centering
\includegraphics[width=0.8\textwidth]{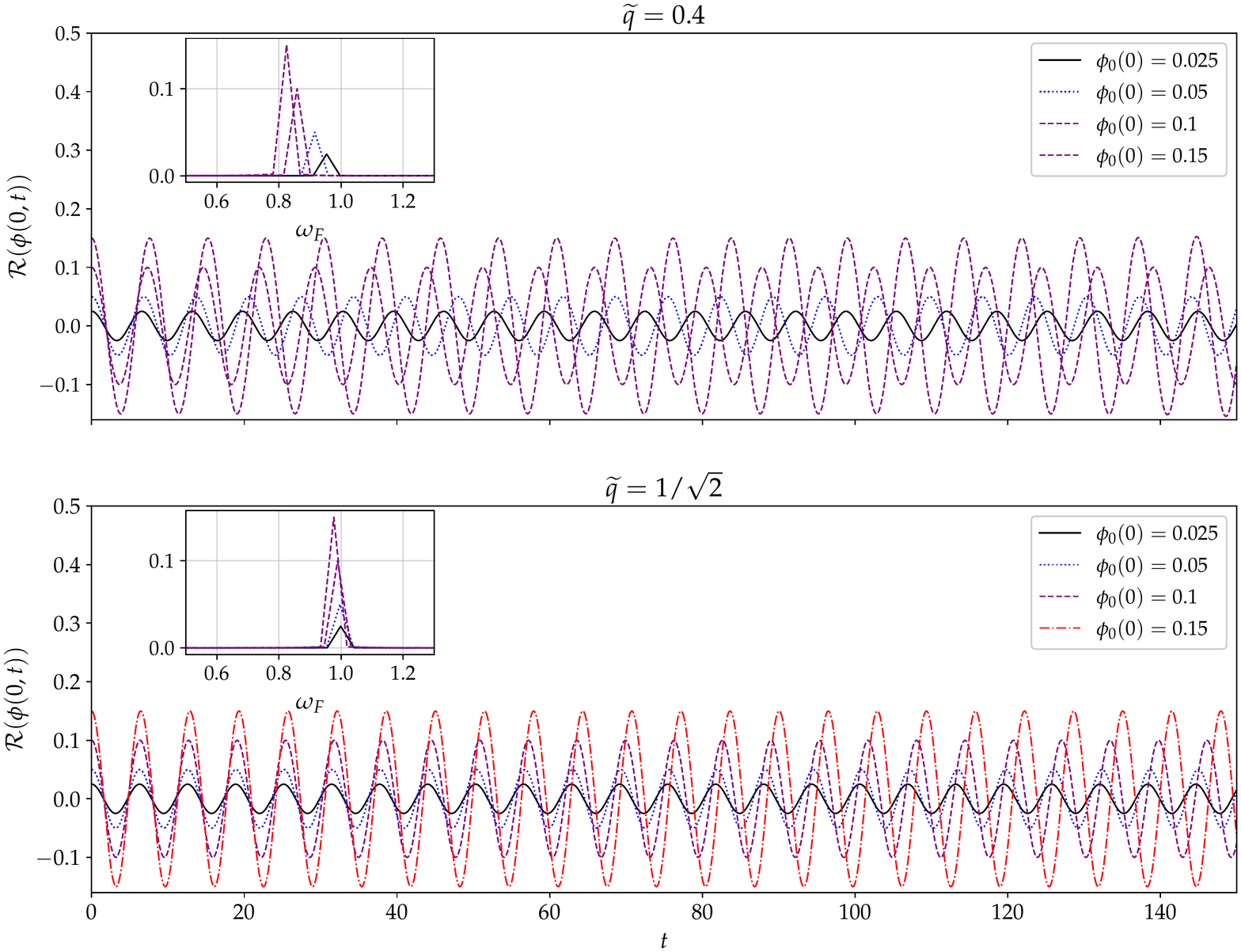}
\caption{Time evolution of the real part of the scalar field evaluated at the origin, $Re(\phi(t,r=0))$, for the different boson star models described in the previous Section.  {\em Top panel:} Models with charge $\tilde{q}=0.4$ corresponding to Figure~\ref{fig:field_0_4}. {\em Bottom panel:} Models  with charge $\tilde{q}=1/\sqrt{2}$ corresponding to Figure~\ref{fig:field_0_crit}. The inset shows the frequencies obtained by performing a Fourier transform of these time evolution data.}
\label{fig:oscilations_t=150}
\end{figure}

\begin{figure}[ht]
\centering
\includegraphics[width=0.8\textwidth]{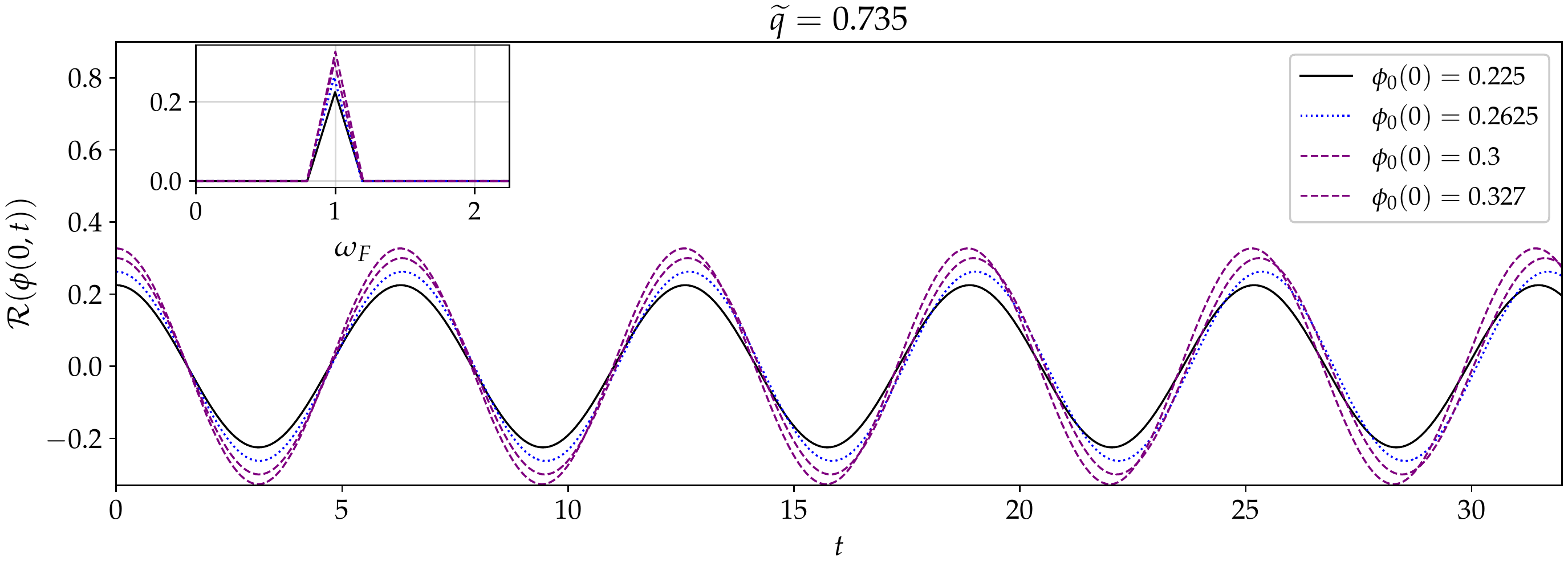}
\caption{Time evolution of the real part of the scalar field evaluated at the origin, $Re(\phi(t,r=0))$, for the different boson star models shown in Figure~\ref{fig:field_0_735} with supercritical charge $\tilde{q}=0.735$. The inset shows the frequencies obtained by performing a Fourier transform of these time evolution data.}
\label{fig:oscilations_t=24}
\end{figure}

The numerical solutions for charged boson stars can be taken as initial data for dynamical evolutions. We have performed short evolutions of the different configurations described in the previous section using the OllinSphere code, which is a fully non-linear time evolution code for numerical relativity in spherical symmetry previously described in~\cite{Torres_2014,Alcubierre:2010ea,Ruiz:2012jt}.  We evolve the unperturbed data in order to verify that the different frequencies obtained in our analysis do correspond to the frequencies observed during dynamical evolution (we will leave a detailed study of the evolution of perturbed solutions for a future work). The numerical evolution code integrates the Einstein equations in time, using the Baumgarte--Shapiro--Shibata--Nakamura formulation adapted to spherical symmetry~\cite{Shibata95,Baumgarte:1998te,Brown:2009dd,Alcubierre:2010is}, coupled with the Klein--Gordon equation for the complex scalar field, and the Maxwell equations for the electromagnetic field as described in~\cite{Torres_2014}. 

Figure~\ref{fig:oscilations_t=150} shows the time evolution of the (unperturbed) real part of the scalar field evaluated at the origin, $Re(\phi(t,r=0))$, for two of the boson star families described in the previous section. In particular, we consider the families with charges $\tilde{q}=0.4$ and $\tilde{q}=1/\sqrt{2}$, corresponding to Figures~\ref{fig:field_0_4} and~\ref{fig:field_0_crit}. The evolutions presented here were performed with a resolutions of $\Delta r = 0.005$ and a Courant parameter $\Delta t/\Delta r=0.5$. 

Similarly, Figure~\ref{fig:oscilations_t=24} shows the time evolution of the (unperturbed) real part of the scalar field evaluated at the origin for the boson star models corresponding to Figure~\ref{fig:field_0_735}, which are solutions with a super-critical charge $\tilde{q}=0.735$. As discussed above, these super-critical solutions are all gravitationally unbound and are therefore expected to be unstable. Indeed, in our numerical simulations we have found that even a small discretization error is sufficient to trigger either the collapse of these solutions to a black hole, or else their dispersion to infinity, in a relatively short time (we will discuss this in detail in a future publication).  Because of this, in order to reduce the discretization error, for these evolutions we have increased the resolution so that we now take $\Delta r = 0.001$ (with the same Courant parameter as before). Also, in the Figure we only plot the time evolutions up to a time $t \sim 32$ (whereas the time evolutions of Figure~\ref{fig:oscilations_t=150} are plotted to much larger times).


\subsection{Summary of properties of our particular configurations}

Finally, in Table~\ref{tab:frequencies} we show the frequencies $\omega$ obtained for each of these specific configurations discussed above (after the rescaling and gauge transformation described in Section~\ref{sec:boundaries}), as well as the total charge $Q$, total mass $M$, total particle number $N$, effective radius $R_{99}$, and binding energy $E_B$.  Do notice that the table shows the total charge $Q = qN = \sqrt{2} \: \tilde{q} N$, and not the rescaled charge $\tilde{Q}$, this is in order to make it easier to compare $Q$ directly with the total mass $M$. From the table one can see that in all cases we have $Q<M$, with the particular configuration with critical charge $\tilde{q}=1/\sqrt{2}$ and small amplitude $\phi_0(0)=0.025$ having $Q$ very close to $M$, though still smaller.

The table also includes the frequencies of oscillation $\omega_{F}$ obtained from a Fourier transform of the time evolution data from Figures~\ref{fig:oscilations_t=150} and~\ref{fig:oscilations_t=24} (shown in the insets of those Figures). One can see that the frequencies obtained from the solution of the eigenvalue problem do indeed correspond, to several decimal places, with the frequencies obtained from the Fourier transform of the time evolution.

\begin{table}
\begin{tabular}{c c c c c c c c c}
\hline
$\phi_0(0)$ & $\omega$  & $\omega_{F}$ & $Q$ & $M$      & $N$      & $R_{99}$ & $E_B$     &  $\tilde{q}$ \\
\hline \hline
0.025    & 0.95377   & 0.95344  &  0.38430  & 0.67018  & 0.67936  & 16.57183 & -0.00917  &  0.4 \\
0.05     & 0.91539   & 0.91512  &  0.45234  & 0.78303  & 0.79964  & 10.97510 & -0.01661  &  0.4 \\
0.1      & 0.85837   & 0.85821  &  0.44575  & 0.77361  & 0.78798  &  6.97835 & -0.01436  &  0.4 \\
0.15     & 0.82490   & 0.82447  &  0.37916  & 0.67488  & 0.67027  &  5.32124 & 0.00461   &  0.4 \\
\hline
0.025    & 0.99922   & 0.99888  &  11.45086  & 11.45981 & 11.45086 & 46.11791 & 0.00894   & $1/\sqrt{2}$ \\
0.05     & 0.99693   & 0.99657  &  5.65906  & 5.67695  & 5.65906  & 23.22937 & 0.01788   & $1/\sqrt{2}$ \\
0.1      & 0.98842   & 0.98876  &  2.69844  & 2.73410  & 2.69844  & 11.99448 & 0.03565   & $1/\sqrt{2}$ \\
0.15     & 0.97666   & 0.97657  &  1.65972  & 1.71259  & 1.65972  & 8.53394  & 0.05287   & $1/\sqrt{2}$ \\
\hline
0.225    & 0.99795   & 0.99796  &  1.46366  & 1.49587  & 1.40811  & 17.84374 & 0.08776   &  0.735 \\
0.2625   & 0.99122   & 0.99119  &  1.08345  & 1.13212  & 1.04233  & 12.65534 & 0.08979   &  0.735 \\
0.3      & 0.99358   & 0.99354  &  1.06661  & 1.11613  & 1.02614  & 16.22238 & 0.08999   &  0.735 \\
0.327    & 0.99996   & 1.00002  &  1.54062  & 1.57106  & 1.48216  & 48.33719 & 0.08890   &  0.735 \\
\hline
\end{tabular}
\caption{Charged boson stars models for different values of the charge $\tilde{q}$ and different central amplitudes $\phi_0(0)$. We show the frequency obtained from the solution of the eigenvalue problem $\omega$ (after the rescaling and gauge transformation described in the text), as well as the total charge $Q$, total mass $M$, total particle number $N$, effective radius $R_{99}$, and binding energy $E_B$.  We also show the frequencies obtained from a Fourier transform of the dynamical evolution of the scalar field $\omega_{F}$.}
\label{tab:frequencies}
\end{table}


\section{DISCUSSION AND CONCLUSIONS}
\label{sec:CONCLUSIONS}

In this paper we have considered solutions to the spherically symmetric stationary EMKG system commonly known as ``charged boson stars'', previously studied in~\cite{Jetzer89,Pugliese_2013,Kain_2021}. We have presented solutions corresponding to the ground state (i.e. with no nodes on the scalar field), for both sub-critical charges with $\tilde{q} \leq 1/\sqrt{2}$ and super-critical charges with $\tilde{q} > 1/\sqrt{2}$, where $\tilde{q}$ is defined in terms of the charge and mass parameters $q$ and $m$ of the scalar field as $\tilde{q}= (q/m)/\sqrt{2}$.

For the sub-critical cases with $\tilde{q} \leq 1/\sqrt{2}$, we find that solutions can in fact exist for all possible values of the central scalar field $\phi_0(0)$. In particular, for the behavior of the frequency $\omega$ we find that boson stars with a sub-critical charge have $\left. (d\omega/d\phi_0) \right|_{\phi_0(0)=0}<0$, while for boson stars with precisely the critical charge $\tilde{q} = 1/\sqrt{2}$ we find instead $\left. (d\omega/d\phi_0) \right|_{\phi_0(0)=0} \simeq 0$. Furthermore, we have shown that is possible to find solutions for configurations with super-critical charges $\tilde{q} > 1/\sqrt{2}$, but only for a finite range of values of $\phi_0(0)$, and a limited range of the charge $\tilde{q}$ such that $1/\sqrt{2} < \tilde{q} \lesssim 0.739$. Outside this range the frequency of the field becomes greater than the boson mass parameter, $\omega>m$, which implies that exponentially decreasing solutions cannot exist.  In particular, we find that is not possible to find super-critical solutions for very small central amplitudes $\phi_0(0) \approx 0$, and as the boson charge increases the allowed range of values for $\phi_0(0)$ becomes narrower, so that for $\tilde{q} \gtrsim 0.739$ no more solutions are found.

Boson stars with a sub-critical charge $\tilde{q} \leq 1/\sqrt{2}$ are similar to standard $\tilde{q}=0$ boson stars in the sense that they have a local maximum of the total mass $M$ for a finite value of $\phi_0(0)$, although for configurations with the critical charge this local maximum moves to $\phi_0(0) \approx 0$ while the maximum mass seems to diverge to infinity, $M \rightarrow \infty$. On the other hand, for super-critical charges $1/\sqrt{2} \leq \tilde{q} < 0.739$ this behavior changes and there is no longer a local maximum of the mass. Another interesting feature is that for configurations with $\tilde{q} \gtrsim 0.7$ the total mass $M$ is always larger than the maximum allowed mass for a $\tilde{q}=0$ boson star $M_\mathrm{Kaup}$.

For charged boson stars with a critical charge $\tilde{q}=1/\sqrt{2}$ we have also found that their compactness defined as $C=M/R_{99}$ behaves as an upper limit for the compactness all charged boson star configurations, both sub-critical and super-critical. In particular, we find that for boson stars with the critical charge, even if the total mass $M$ and total charge $Q$ seem to diverge to infinity for $\phi_0(0) \approx 0$, the compactness approaches a maximum value of $C_{99} \approx 0.25$. On other hand, for super-critical solutions with $\tilde{q}>1/\sqrt{2}$ the compactness falls again to smaller values.

As we have argued above, the disappearance of the local maximum of the total mass $M$ for super-critical configurations would seem to indicate that all such solutions are unstable.  This conclusion is strengthened by considering the binding energy $E_B$ for the different configurations. Indeed, we find that for boson stars with a sub-critical charge there always exists a region where the solutions are gravitationally bound such that their binding energy is negative, $E_B<0$. On the other hand, all solutions with super-critical charges have a positive binding energy $E_B>0$, so they are gravitationally unbound. In relation to this, one can show that the binding energy is directly related to the total charge to mass ratio $Q/M$ as $E_B = M [1 - (m/q)(Q/M)] = M(1 - \tilde{Q}/\tilde{q})$, where $\tilde{Q}$ is now defined as $\tilde{Q} = (Q/M)/\sqrt{2}$. From this we find that all super-critical solutions are such that $\tilde{Q} < \tilde{q}$.

But from our results we in fact find the much stronger conclusion that {\em all}\/ charged boson star configurations, both sub-critical and super-critical, are such that $Q \leq M$ ($\tilde{Q} \leq 1/\sqrt{2}$), with the equality only achieved for the specific case of a critical charge $\tilde{q}=1/\sqrt{2}$ when $\phi_0(0) \rightarrow 0$. This implies that, even though we do find a family of super-critical solutions in the sense that $q>m$ ($\tilde{q}>1/\sqrt{2}$), there are in fact {\em no}\/ super-critical solutions in the sense that $Q>M$. In other words, in general relativity it is not the ratio between the charge and mass parameters $q$ and $m$ the one that determines the existence of solutions, but rather the ratio between the total charge $Q$ and the total mass $M$, as one would have expected on physical grounds.

Finally, we have also performed some preliminary time evolutions for unperturbed sub-critical and super-critical configurations in order to verify that the different frequencies obtained from our solution to the eigenvalue problem in fact correspond to the frequencies observed during a dynamical evolution, and find that the frequencies do coincide to several decimal places, indicating that our solution to the eigenvalue problem is correct.  Our numerical evolutions also indicate that super-critical configurations are indeed unstable, as even a (small) numerical discretization error is enough to trigger either collapse to a black hole, or dispersion to infinity (though we will report on this in detail elsewhere).


\bibliographystyle{apsrev4-2}
\bibliography{referencias}


\end{document}